\documentclass[conference]{IEEEtran}
\IEEEoverridecommandlockouts
\usepackage{amssymb}
\usepackage{cite}
\usepackage{graphicx}
\usepackage{array}
\usepackage{multirow}
\usepackage{amsmath}
\usepackage{stfloats}
\usepackage{graphicx}
\usepackage{tabularx}
\usepackage{epsfig,epsf}
\usepackage{setspace}
\usepackage{bm}
\usepackage{textcomp}
\usepackage{algorithmic}
\usepackage{algorithm}
\usepackage{graphicx}
\usepackage{setspace}
\usepackage{url}
\usepackage{verbatim}
\usepackage{graphicx}
\usepackage{subcaption}
\usepackage{amssymb}
\usepackage{color}
\usepackage{xpatch}
\definecolor{shadecolor}{RGB}{0,0,255}
\definecolor{blue}{RGB}{0,0,255}

\newtheorem{theorem}{Theorem}

\newtheorem{lemma}{Lemma}

\newtheorem{remark}{Remark}

\makeatletter

\ExplSyntaxOn
\cs_new:Npn \bibColoredItems #1#2
{
	\clist_map_inline:nn {#2} { \cs_new:cpn {bib@colored@##1} {#1} } 
}
\ExplSyntaxOff

\newcommand\bib@setcolor[1]{%
	\ifcsname bib@colored@#1\endcsname
	\expanded{\noexpand\color{\csname bib@colored@#1\endcsname}}%
	\else
	\normalcolor
	\fi
}

\xpatchcmd\@bibitem
{\item}
{\bib@setcolor{#1}\item}
{}{\fail}

\xpatchcmd\@lbibitem
{\item}
{\bib@setcolor{#2}\item}
{}{\fail}
\makeatother

\begin{document}
	
\title{Signal-Dependent Shot Noise Modeling of Rydberg Atomic Quantum Receivers: A Design Perspective}

\author{Qihao Peng, Qu Luo,~\IEEEmembership{Member,~IEEE}, Tierui Gong,~\IEEEmembership{Member,~IEEE}, Neng Ye,~\IEEEmembership{Senior Member,~IEEE}, 
Jizhou Wu, \\Cunhua Pan,~\IEEEmembership{Senior Member,~IEEE}, Maged Elkashlan,~\IEEEmembership{Senior Member,~IEEE}, Pei Xiao,~\IEEEmembership{Senior Member,~IEEE}, \\Chau Yuen,~\IEEEmembership{Fellow,~IEEE}, George K. Karagiannidis,~\IEEEmembership{Fellow,~IEEE}, Jiangzhou Wang,~\IEEEmembership{Fellow,~IEEE}.
\vspace{-1cm}
		\thanks{Q. Peng, Q. Luo, and P. Xiao are affiliated with 5G and 6G Innovation Centre, Institute for Communication Systems (ICS) of the University of Surrey, Guildford, GU2 7XH, UK. (e-mail: \{q.peng, q.u.luo, p.xiao\}@surrey.ac.uk).
        }\\
        \thanks{N. Ye is with the School of Cyberspace Science and Technology, Beijing
Institute of Technology, Beijing 100081, China (email: ianye@bit.edu.cn).} \\
\thanks{J. Wu is with State Key Laboratory of Quantum Optics Technologies and Devices, Institute of Laser Spectroscopy, Shanxi University, Taiyuan 030006, China (e-mail:  wujz@sxu.edu.cn).}
         \thanks{C. Pan and J. Wang are with the National Mobile Communications Research Laboratory, Southeast University, Nanjing, China. (e-mail: cpan@seu.edu.cn;j.z.wang@seu.edu.cn).} \\
             \thanks{T. Gong and C. Yuen are with the School of Electrical and Electronics Engineering, Nanyang Technological University, Singapore 639798 (e-mail: trgTerry1113@gmail.com, chau.yuen@ntu.edu.sg). }
          \thanks{M. Elkashlan is with the School of Electronic Engineering and Computer Science, Queen Mary University of London, E1 4NS London, U.K.(e-mail: maged.elkashlan@qmul.ac.uk).} \\ 
             \thanks{George K. Karagiannidis is with the Department of Electrical and Computer Engineering, Aristotle University of Thessaloniki, 541 24 Thessaloniki, Greece (e-mail: geokarag@auth.gr).} 
    
} 
\maketitle

\begin{abstract}
    In this paper, \textcolor{black}{we develop a communication-oriented complex baseband equivalent model for superheterodyne Rydberg atomic quantum receivers (RAQRs). The model explicitly captures photodetection-induced signal-dependent shot noise and its coupling with the optical operating point.} By leveraging an atomic superheterodyne architecture and a strong local oscillator, we construct a complex baseband representation for both the received signal and the signal-dependent shot noise under both direct incoherent optical detection and balanced coherent optical detection. The derived model reveals that the optical operating point jointly determines the normalized effective receive gain and the equivalent noise background, thereby establishing a traceable gain-noise tradeoff governed by system design. More importantly, the proposed model shows that neglecting signal-dependent shot noise may lead to inaccurate operating-point design. Finally, by extending to the multiple-input-multiple-output (MIMO) case, we derive a lower bound on the achievable rate while considering the signal-dependent shot noise. \textcolor{black}{Our analysis \textcolor{black}{reveals} that the non-zero asymptotic rate of RAQ-MIMO and its superiority over conventional RF-MIMO hinge on the normalized noise floor of the RAQ receive chain falling below that of RF MIMO.} Simulation results validate our analysis and yield practical, closed-form design guidelines for RAQR front ends, revealing parameter regimes in which RAQ-MIMO outperforms conventional MIMO systems.

\end{abstract}	
	

\section{Introduction}

Future wireless receivers call for high sensitivity, broad spectral tunability, and compact front ends, which has motivated growing interest in Rydberg atomic quantum receivers (RAQRs) as a fundamentally different alternative to conventional electronic radio-frequency (RF) front ends  \cite{artusio2022modern}. By exploiting the interaction between incident RF fields and atomic energy levels, RAQRs transduce RF signals into measurable optical responses through probe-beam readout, thereby enabling a physically interpretable RF-to-optical reception mechanism  \cite{tanasittikosol2011microwave}. Owing to this unique transduction process, RAQRs are emerging as a promising receiver architecture for future wireless communication networks \cite{QihaoWCM,chen2025new}.


Early studies established the physical foundation of Rydberg atomic reception by demonstrating electromagnetically induced transparency (EIT)-based optical readout of highly excited Rydberg states and microwave electrometry via EIT/Autler–Townes splitting (ATS) \cite{PhysRevLett,sedlacek2012microwave}. Subsequent experiments characterized the impacts of the RF signal on optical readout, validating these foundations \cite{fan2015atom,fan2015effect}. Building on these mechanisms, follow-up work explored the use of Rydberg atoms as wireless receivers. In particular, by monitoring the RF-induced variation of the EIT/ATS response, Rydberg atom-based receivers were designed for amplitude modulation \cite{meyer2018digital,li2022rydberg}, frequency modulation \cite{song2019rydberg,holloway2020multiple}, and phase modulation \cite{holloway2019detecting,cai2023high}. In addition, it has been shown that the sensitivity of RAQRs has reached the order of \(\mu{\rm V}/\text{cm}/\sqrt{\text{Hz}}\) by using the four-level \textcolor{black}{ladder-type} scheme \cite{fancher2021rydberg,liu2022highly,liu2023electric}. The sensitivity of RAQRs has been experimentally shown to the level of \({\rm nV}/\text{cm}/\sqrt{\text{Hz}}\) by leveraging the superheterodyne architecture \cite{simons2019rydberg,jing2020atomic}. These experimental advances mainly demonstrate the reception feasibility and sensitivity potential of RAQRs, while their communication-level equivalent modeling remains insufficiently understood.

From a communication-theoretic perspective, recent efforts have started to bridge Rydberg quantum sensing and wireless system design. In particular, Cui \emph{et al.} proposed atomic multiple-input-multiple-output (MIMO) receivers and showed that signal detection is intrinsically a nonlinear biased phase-retrieval problem, rather than the linear Gaussian model encountered in conventional MIMO systems \cite{cui2025towards}. Yuan \emph{et al.} further developed an electromagnetic modeling framework for Rydberg atom-based MIMO systems and characterized their spatial multiplexing capability from a propagation and capacity perspective \cite{Yuanmodel}. However, the above magnitude-based model imposed scalability limitations due to phase ambiguity. To alleviate the scalability bottleneck, a reconfigurable intelligent surface (RIS) was introduced for amplitude modulation in \cite{peng2025risassistedatomicmimoreceiver}. Beyond magnitude-only reception, Gong \emph{et al.} developed an equivalent complex model for time-varying signal reception, leveraging the steady-state quantum transitions \cite{gong2026rydbergatomicquantumreceivers}. Furthermore, they extended it to the multi-user uplink and derived equivalent Rydberg atomic quantum (RAQ)-MIMO baseband models and achievable-rate lower bounds under maximum ratio combining (MRC) and zero-forcing (ZF) schemes \cite{gong2025rydbergatomicquantummimo}. More recently, the authors of \cite{11278503} moved beyond steady-state abstractions by introducing transfer-function-based and general dynamic signal models, thereby capturing the response of Rydberg receivers to time-varying signals. Related extensions have also emerged for channel estimation \cite{kim2025quantum,xu2025channel}, direction of arrival estimation \cite{gong2025rydberg}, and localization \cite{location,LEOLocalization}. Despite these advances, most existing works focus on signal representation, channel modeling, or sensing functionality, while the communication impact of photodetection-induced signal-dependent noise remains largely absent.

Existing studies on RAQR's noise have already shown that performance is fundamentally constrained by noise floor. Specifically, \textcolor{black}{Wang \emph{et al.} \cite{wang2023noise} characterized the noise power spectral density in the optical-intensity domain, but did not develop a complex baseband model.  Tang \emph{et al.} further studied the noise performance of RAQRs by explicitly accounting for quantum projection noise but treated shot noise as additive, not signal-dependent \cite{tang2025noise}.}  Meanwhile, although recent studies \cite{cui2025towards,gong2026rydbergatomicquantumreceivers,11278503} have developed communication-oriented equivalent signal models for RAQRs, an end-to-end communication-level complex baseband model that explicitly captures the photodetection-induced noise and its coupling with optical parameters has remained largely unexplored.

\textcolor{black}{Our paper builds directly on \cite{gong2026rydbergatomicquantumreceivers} and fills this specific gap by incorporating the missing signal-dependent shot noise into the complex baseband model and analyzing the gain–noise tradeoff, the optimal optical operating point, and the RAQ-MIMO achievable rate.} Our contributions are summarized as follows:
\begin{enumerate}

\item Instead of assuming a priori signal-independent additive disturbance, we derive an equivalent baseband model for signal-dependent shot noise based on superheterodyne architecture. Leveraging standard baseband signal processing, the real-valued optical intensity is further transformed into the complex-valued domain, yielding a tractable equivalent complex baseband model for signal-dependent shot noise. Simulation results based on the master equation verify the accuracy of our derived model of signal-dependent shot noise.

\item We characterize the composite noise structure of RAQRs by jointly accounting for user-signal-dependent shot noise, direct-current-dependent shot noise, and thermal noise. Based on the derived gain–noise relationship, we obtain closed-form design criteria for maximizing the effective received signal-to-noise ratio (SNR), thereby providing explicit guidelines for practical optical front-end design. Additionally, simulation results validate our analysis and demonstrate that the sensitivity of RAQRs is more strongly influenced by the local oscillator (LO) and the probe beam.

\item By extending the single-input-single-output (SISO) case to a general multiuser RAQ-MIMO uplink model, we establish a new system model that captures the dependence on user signals. Based on this model, we derive lower bounds on the achievable rates under MRC and ZF detection, thereby revealing the impact of optical front-end parameters on system performance. Furthermore, simulation results validate the accuracy of the derived bounds and demonstrate that linear detection remains effective for RAQ-MIMO while preserving the classical power-scaling law.

\item \textcolor{black}{By comparing RAQ-MIMO with conventional RF-MIMO under the same array size and fading conditions, we identify an explicit crossover condition that is governed primarily by the LO power and the probe-beam power. Once the power of LO and probe beam exceed their critical levels, i.e.,  \(P_\text{LO}^*\) or \(P_\text{0}^*\), the advantage of RAQ-MIMO over conventional MIMO diminishes and may even vanish. This result shows that front-end design is the dominant factor for realizing the promise of atomic reception.} 

\end{enumerate}

\emph{Organization and Notations}: The remaining sections are organized as follows. The principles of quantum measurement on a weak RF signal are presented in Section II. The equivalent models of RF-to-optical conversion and shot noise are derived in Section III. In Section IV, we extend the model into a general atomic array and provide a closed-form expression for the lower bound of the data rate. The simulated results are presented in Section V. Finally, the conclusions are drawn in Section VI. The notations
are given in the following: \(\hbar\) is the reduced Planck constant and \(j^2 = -1\). \(\epsilon_0\) and \(c\) denote the vacuum permittivity and speed of light, respectively.  \(q\) and \(a_0\) respectively represent the elementary charge and Bohr radius. \(\Re (\cdot)\) and \(\Im(\cdot)\) denote the real and imaginary parts of \((\cdot)\), respectively. \(\eta_1\) is the quantum efficiency of the photodetector.

\section{Quantum Response Model}
In this section, we will briefly introduce the quantum response of a weak electromagnetic (EM) signal in a four-level atomic system.

\subsection{Principles of Quantum Measurement}
\begin{figure*}
     \centering
        \includegraphics[width=0.95\linewidth]{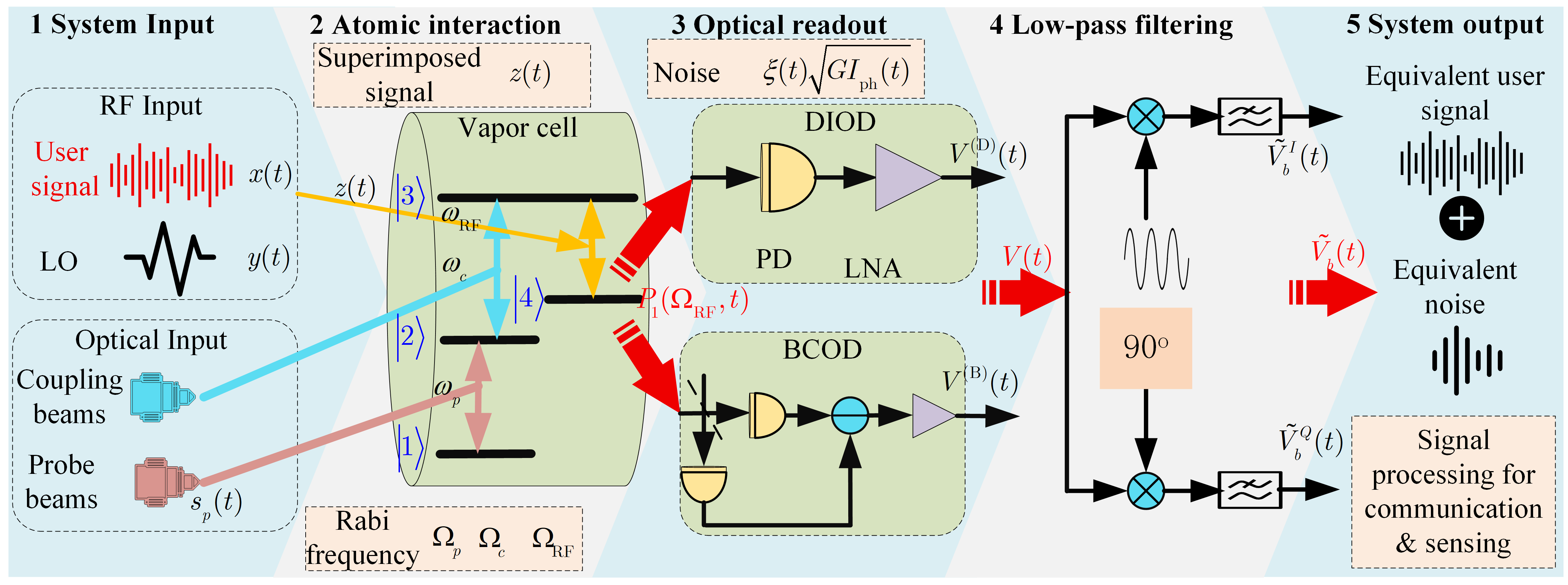}
        \caption{\textcolor{black}{4-level superheterodyne architecture of RAQR.}}
        \label{systemmodel}
\end{figure*}

In Fig.~\ref{systemmodel}, we consider a four-level ladder (or near-ladder) atomic configuration with states $\{|1\rangle,|2\rangle,|3\rangle,|4\rangle\}$, where optical fields address the transitions of $|1\rangle\rightarrow|2\rangle$ and $|2\rangle\rightarrow|3\rangle$ and the weak EM signal drives the transition of $|3\rangle\rightarrow|4\rangle$. Specifically, (i) a probe laser couples $|1\rangle\rightarrow|2\rangle$ with a Rabi frequency $\Omega_p$ and the dipole moment \(\mu_{12}\), (ii) a coupling laser couples $|2\rangle\rightarrow|3\rangle$ with Rabi frequency $\Omega_c$ and the dipole moment \(\mu_{23}\), and (iii) the weak EM field (radio-frequency/LO signal) couples $|3\rangle\rightarrow|4\rangle$ with Rabi frequency $\Omega_{\mathrm{RF}}$ and the dipole moment \(\mu_{34}\). Furthermore, the corresponding detunings of probe beam, coupling beam, and RF signal are defined as $\Delta_p$, $\Delta_c$, and $\Delta_{\mathrm{RF}}$, respectively.

Defining $\boldsymbol{\rho}$ as the density matrix, the dynamics of the four-level transition scheme can be characterized by 

\begin{equation}
\begin{split}
    \frac{\text{d}\boldsymbol{\rho} }{t} &= -j [\mathbf{H},\boldsymbol{\rho}] - \frac{1}{2}\{\boldsymbol{\Gamma},\boldsymbol{\rho} \} + \boldsymbol{\Lambda},\\
    \mathbf{H} &= \left[\begin{array}{cccc}
         0 & \frac{\Omega_p}{2} & 0& 0\\
          \frac{\Omega_p}{2}& \Delta_p & \frac{\Omega_c}{2} & 0\\
         0 &  \frac{\Omega_c}{2}& \Delta_p + \Delta_c &  \frac{\Omega_\text{RF}}{2} \\
         0& 0&\frac{\Omega_\text{RF}}{2} &\Delta_p + \Delta_c + \Delta_\text{RF}
    \end{array}\right], \\
    \boldsymbol{\Gamma} &= \text{diag}\{\gamma,\gamma+\gamma_2,\gamma+\gamma_3+\gamma_c,\gamma+\gamma_4\},\\
    \boldsymbol{\Lambda}& =  \text{diag}\{\gamma+\gamma_2\rho_{22}+\gamma_4\rho_{44},\gamma_{3}\rho_{33},0,0\},
\end{split}
\end{equation}
where \(\mathbf{H}\), \(\boldsymbol{\Gamma}\), and \(\boldsymbol{\Lambda}\) represent the Hamiltonian, the relaxation matrix, and the decay matrix, respectively. \(\rho_{mn}\) represents the \([m,n]\)-th element of \(\boldsymbol{\rho}\), \(\gamma_i\), \(i \in \{1,2,3,4\}\), is the spontaneous decay rate in the \(i\)-th level, \(\gamma\) and \(\gamma_c\) mean the relaxation rates relying on the atomic transition and collision, respectively. \textcolor{black}{For analytical tractability, we neglect the collision-induced dephasing and the decay rates of the two upper Rydberg states, i.e., $\gamma_c=0$ and $\gamma_3=\gamma_4=0$  \cite{jing2020atomic}. This is not a claim that these processes are absent, but a statement that their rates are much smaller than the intermediate-state decay, the Rabi couplings, and the resulting EIT/ATS linewidth. \footnote{This regime is standard in Rydberg EIT experiments with dilute, buffer-gas-free vapor cells and frequency-stabilized narrow-linewidth lasers, where the upper Rydberg states are long-lived and collisional dephasing is weak. Under these conditions, $\gamma_c$, $\gamma_3$, and \(\gamma_4\) contribute only minor linewidth broadening. Therefore, neglecting their impacts is a standard approximation for deriving a compact closed-form RF-to-optical transfer function.}}

Based on the above discussions, we set \( \frac{\text{d}\boldsymbol{\rho} }{\text{d}t}  = 0\) and obtain the steady-state solution of density matrix \(\boldsymbol{\rho}\). Then, we are interested in \(\rho_{21}\) of  \(\boldsymbol{\rho}\), as the measured susceptibility based on probe beam is proportional to $\rho_{21}$, which is given by \cite{gong2026rydbergatomicquantumreceivers}
\begin{equation}
    \begin{split}
 & \rho_{21}(\Omega_{\mathrm{RF}})=\Omega_{p}\times \notag \\
 & \frac{A_{1}\Omega_{\mathrm{RF}}^{4}+A_{2}\Omega_{\mathrm{RF}}^{2}+A_{3}-j\left(B_{1}\Omega_{\mathrm{RF}}^{4}+B_{2}\Omega_{\mathrm{RF}}^{2}+B_{3}\right)}{C_{1}\Omega_{\mathrm{RF}}^{4}+C_{2}\Omega_{\mathrm{RF}}^{2}+C_{3}},
    \end{split}
\end{equation}
where \(A_1\), \(A_2\), \(A_3\), \(B_1\), \(B_2\), \(B_3\), \(C_1\), \(C_2\), \(C_3\) can be found in Appendix A of \cite{gong2026rydbergatomicquantumreceivers}.

\subsection{RF-Optical System Model}
As previously described, the RAQR can perform RF-Optical conversion by using probe-based susceptibility. We assume that the probe beam at the access area of the atomic vapor cell is written as 
\begin{equation}
    s_p(t) = {U_0}\cos(2\pi f_pt + \phi_0),
\end{equation}
where \textcolor{black}{\(U_0\), \( f_p\), \(\phi_0\) are the electric field, frequency, and phase of the input probe beam, respectively. \(P_0 = \frac{\pi c\epsilon_0}{8\ln 2}F_p^2\left|U_0\right|^2\) denotes the power of the input beam, where \(F_p\) is the full width at half maximum (FWHM) of the probe beam.} After passing through the atomic vapor cell, the output probe is given by
\begin{equation}\label{ouputprobe}
    s_p(\Omega_\text{RF},t) = {U_p(\Omega_{\mathrm{RF}})}\cos\Big[2\pi f_pt+\phi_p(\Omega_\text{RF})\Big],
\end{equation}
where  \(U_p(\Omega_\text{RF})\) and phase \(\phi_p(\Omega_\text{RF})\) are the amplitude and phase related to the RF signal, which are given by
\begin{equation}
    \begin{split}
        U_p(\Omega_{\mathrm{RF}}) &= U_0 \exp\!\left(-\frac{\pi d}{\lambda_p}\Im\{\chi(\Omega_{\mathrm{RF}})\}\right), \\
\phi_p(\Omega_{\mathrm{RF}}) &= \phi_0 + \frac{\pi d}{\lambda_p}\Re\{\chi(\Omega_{\mathrm{RF}})\}.
    \end{split}
\end{equation}
 \(\lambda_p\) and \(d\) are the wavelength of the probe beam and the length of the vapor cell, respectively.  The output power of the probe beam is \(P_1(\Omega_\text{RF}) = \frac{\pi c\epsilon_0}{8\ln 2}F_p^2\left|U_p(\Omega_\text{RF})\right|^2\). \(\chi(\Omega_{\text{RF}})\) is the probe-based susceptibility, which is given by
\begin{equation}
      \chi(\Omega_{\text{RF}}) = -\frac{2N_0 \mu^2_{12}}{\epsilon_0\hbar \Omega_p}\rho_{21}(\Omega_{\text{RF}}),
\end{equation}
where \(N_0\) is the atomic density. 

\begin{figure}[t]
    \centering
    \includegraphics[width=0.8\linewidth]{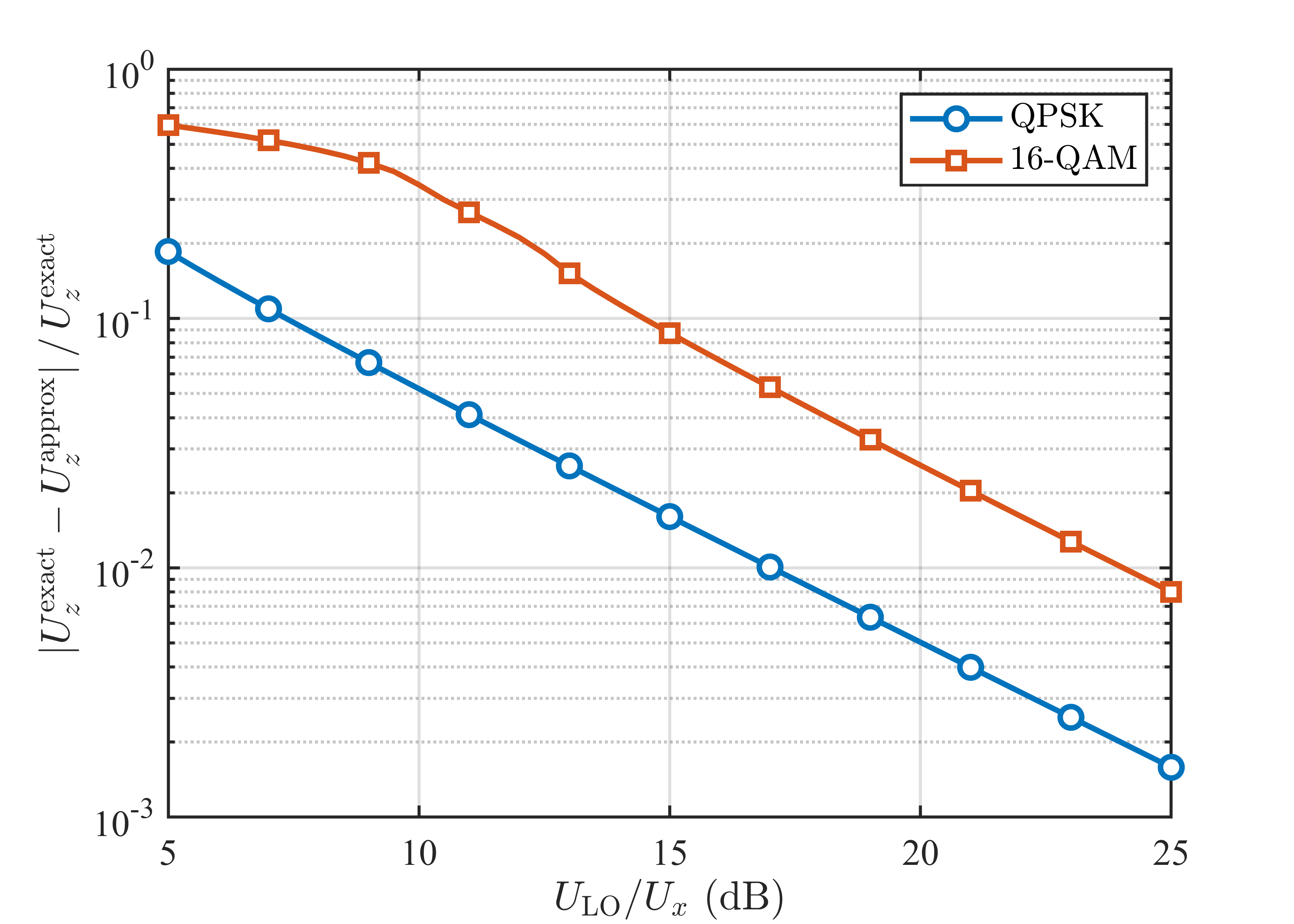}
    \caption{\textcolor{black}{Approximation error under various ratios of electrical field.}}
    \label{ULOUX}
\end{figure}

In the following, we reveal the relationship between the superimposed signal (i.e., weak user's signal and LO signals) and the Rabi frequency \(\Omega_\text{RF}\)\footnote{The detailed process can be found in Section III.A of \cite{gong2026rydbergatomicquantumreceivers}}. By assuming that the weak signal from the user is a plane wave, which is given by
\begin{equation}
    x(t) = {U_x}\cos(2\pi f_c t + \theta_x)=\Re\{x_b(t)e^{j2\pi f_c t}\},
\end{equation}
where \(U_x\), \(f_c\), and \(\theta_x\) are the electric field, carrier frequency, and phase, respectively. \(x_b(t) = {U_x}e^{j\theta_x}\) denotes the equivalent baseband signal. Its corresponding power is \(P_x = \frac{1}{2}c\epsilon_0A_e |U_x|^2\), where \(A_e\) is the effective RAQR's aperture. Similarly, the LO signal can be expressed as
\begin{equation}
    y(t) = {U_\text{LO} }\cos(2\pi f_\text{LO} t+\theta_\text{LO} ),
\end{equation}
where \(U_\text{LO}\), \(f_\text{LO} \), and \(\theta_\text{LO} \) are the electric field, carrier frequency, and phase of LO signal, respectively. The LO's power is \(P_\text{LO}  =\frac{1}{2}c\epsilon_0A_e |U_\text{LO} |^2 \).
By  defining \(f_\delta = f_c - f_\text{LO} \) and \(\theta_\delta = \theta_x -\theta_\text{LO} \) as the frequency difference and phase difference, the superimposed RF signal at atomic vapor cell is 
\begin{equation}
    z(t) = x(t)+y(t)\overset{(a)}{\approx} {U_z}\cos(2\pi f_\text{LO}t +\theta_\text{LO} ),
\end{equation}
where \(U_z\) is the superimposed signal's amplitude, which is written as
\begin{equation}
    U_z \overset{(b)}{\approx} U_\text{LO}  + U_x \cos(2\pi f_\delta t+ \theta_\delta),
\end{equation}
where \((a)\) and \((b)\) hold when \(U_\text{LO}  \gg U_x\), as illustrated in Fig. \ref{ULOUX} \footnote{Although the modulation scheme induces amplitude variations that elevate the approximation error, this error remains negligible with higher voltage. This condition is readily satisfied in communication scenarios, where the LO is placed in close proximity to the atomic sensors.}.
The power of superimposed signal is \(P_z = \frac{1}{2}c\epsilon_0A_e |U_z|^2\). By using \(\Omega_\text{RF} = \frac{\mu_{34}}{\hbar}U_z\), \(\Omega_\text{LO}  = \frac{\mu_{34}}{\hbar}U_\text{LO} \), and \(\Omega_x = \frac{\mu_{34}}{\hbar}U_x\), we have
\begin{equation}\label{Omega_rf}
    \Omega_\text{RF} \approx \Omega_\text{LO}  + \Omega_x \cos(2 \pi f_\delta t + \theta_\delta).
\end{equation}
Substituting these results into \eqref{ouputprobe}, the input weak signal \(x(t)\) can be transformed into optical output.

\section{RF-to-Optical Model with Signal-Dependent Shot Noise}
In this section, we rigorously derive the complex-baseband equivalent model signal-dependent shot noise. Then, we characterize the impacts of different types of noise on the received performance from a theoretical aspect, thereby providing an explicit form for the system design in terms of received SNR.

\subsection{Optical-Voltage Conversion}
\subsubsection{DIOD}
Based on the output probe beam \(s_p(\Omega_\text{RF},t)\), the photocurrent based on the incident beam power is 
\begin{equation}
I_{\mathrm{ph}}^{(\mathrm{D})}(t) = \frac{\eta_1 q}{2\pi\hbar f_p}P_1(\Omega_{\mathrm{RF}})\triangleq \alpha P_1(\Omega_{\mathrm{RF}}),
\end{equation}
where \(\alpha\) denotes the photodetector responsivity. Then, \textcolor{black}{the photocurrent is converted into a voltage across the load, whose value is assumed constant and thus omitted here, and is then further amplified by a low-noise amplifier (LNA)}. In photodetection, the photocurrent fluctuation due to the Poissonian photon-counting process has a variance equal to the mean photocurrent, yielding a standard deviation proportional to \(\sqrt{I_{\mathrm{ph}}^{(\mathrm{D})}(t) }\). For a large number of photons within bandwidth \(B\), the Poisson distribution is well approximated by a Gaussian of matching mean and variance. By ignoring the independent noise, this process can be mathematically written as \cite{wang2020tight}
\begin{equation}
    \begin{split}
          V^{(\mathrm{D})}(t) &= \sqrt{G} I_{\mathrm{ph}}^{(\mathrm{D})}(t) + \xi(t)\sqrt{GI_{\mathrm{ph}}^{(\mathrm{D})}(t)}\\
          & =\sqrt{G}\alpha  P_1(\Omega_{\mathrm{RF}}) +  \xi(t)\sqrt{G\alpha  P_1(\Omega_{\mathrm{RF}})} ,
    \end{split}
\end{equation}
where \(\xi(t)\sqrt{G\alpha  P_1(\Omega_{\mathrm{RF}})}\) is the noise related to the input signal and \(\xi(t) \sim \mathcal{N}(0,\varsigma^2)\) is the Gaussian random variable
describing the input-dependent noise \cite{gao2016modulation,moser2012capacity}. \textcolor{black}{Here, \(\varsigma^2\) encapsulates the Schottky shot-noise prefactor and the detection bandwidth \(B\), namely, \(\varsigma^2 = 2qB\).}

Then, by using \eqref{Omega_rf}, we have
\begin{equation}
\begin{split}
        V^{(\mathrm{D})}(t)
=&\sqrt{G}\alpha P_1(\Omega_{\mathrm{RF}})
+\xi(t)\sqrt{G\alpha P_1(\Omega_{\mathrm{RF}})} \\
\overset{(c)}{\approx}& \sqrt{G}\alpha P_1(\Omega_\text{LO} ) \Big[1- 2\kappa_1(\Omega_\text{LO})U_x\cos(2\pi f_\delta t + \theta_\delta)\Big] \\
+ &  \xi(t) \sqrt{G\alpha P_1(\Omega_\text{LO} )\Big[1- 2\kappa_1(\Omega_\text{LO})U_x\cos(2\pi f_\delta t + \theta_\delta)\Big]  } \\
\overset{(d)}{\approx} &  \sqrt{G}\alpha P_1(\Omega_\text{LO} ) \Big[1- 2\kappa_1(\Omega_\text{LO})U_x\cos(2\pi f_\delta t + \theta_\delta)\Big] \\
 +  &  \xi(t) \sqrt{G\alpha P_1(\Omega_\text{LO} )  }\Big[1- \kappa_1(\Omega_\text{LO})U_x\cos(2\pi f_\delta t + \theta_\delta)\Big],
    \end{split}
\end{equation}
where \((c)\) is obtained by \(f(x) = f(x_0) + f'(x_0)(x-x_0)\) and \((d)\) yields \(\sqrt{1+x}\approx 1 + \frac{x}{2}\) for \(|x|\ll 1\), respectively. 
 \(\kappa_1(\Omega_\text{LO}) = \frac{\pi d\mu_{34}}{\lambda_p\hbar}\Im\{\chi'(\Omega_{\mathrm{LO}})\}\) is the approximated value related to Rabi frequency \(\Omega_\text{LO}\). \(\Im\{\chi'(\Omega_{\mathrm{LO}})\}\) is given by
\begin{equation}
    \begin{split}
\Im\{\chi^{\prime}(\Omega_{\text{LO}})\}=\frac{4N_{0}\mu_{12}^{2}}{\epsilon_{0}\hbar}\Omega_{\text{LO}}\left[\frac{2B_{1}\Omega_{\text{LO}}^{2}+B_{2}}{C_{1}\Omega_{\text{LO}}^{4}+C_{2}\Omega_{\text{LO}}^{2}+C_{3}}\right.\\
\left.-\frac{(B_{1}\Omega_{\text{LO}}^{4}+B_{2}\Omega_{\text{LO}}^{2}+B_{3})\left(2C_{1}\Omega_{\text{LO}}^{2}+C_{2}\right)}{(C_{1}\Omega_{\text{LO}}^{4}+C_{2}\Omega_{\text{LO}}^{2}+C_{3})^{2}}\right].
    \end{split}
\end{equation}

\subsubsection{BCOD}
In this scheme, a strong local optical beam \(s_l(t) = \sqrt{2P_l}\cos(2\pi f_p t + \phi_l)\) is adopted, where \(P_l = \frac{\pi c\epsilon_0}{8\ln 2}F^2_p |U_l|^2\) is the power related to its amplitude \(U_l\) and \(\phi_l\) represents the phase. Then, the probe beam and the local optical beam are combined to form two distinct optical beams, i.e., \(\frac{1}{\sqrt{2}}\Big[ s_l(t)  - s_p(\Omega_\text{RF},t) \Big]\) and \(\frac{1}{\sqrt{2}}\Big[ s_l(t)  + s_p(\Omega_\text{RF},t) \Big]\). Then, two photodetectors generate the corresponding photocurrents, which are combined to generate an output photocurrent \(I_{\mathrm{ph}}^{(\mathrm{B})}(t)\), denoted as
\begin{equation}
\begin{split}
        I_{\mathrm{ph}}^{(\mathrm{B})}(t)& = \alpha\sqrt{P_lP_1(\Omega_\text{RF})}\Big(e^{j\big[\phi_l-\phi_p(\Omega_\text{RF})\big]}+e^{-j\big[\phi_l-\phi_p(\Omega_\text{RF})\big]}\Big).
\end{split}
\end{equation}
\textcolor{black}{Similar to the DIOD case, the output voltage across the load after LNA amplification can be expressed as}

\begin{equation}
\begin{split}
         &V^{(B)}(t) = \sqrt{G}I_{\mathrm{ph}}^{(\mathrm{B})}(t) +  \xi(t)\sqrt{G\alpha\big[P_l + P_1(\Omega_\text{RF})\big]}  \\
   =  & 2\alpha \sqrt{G{P}_l{P}_1(\Omega_{\text{RF}})}\cos\Big[\phi_l - \phi_p(\Omega_\text{RF}) \Big] \\
   +  &\frac{\xi_{+}(t)-\xi_{-}(t)}{\sqrt{2}}\sqrt{G\alpha\big[P_l + P_1(\Omega_\text{RF})\big]} \\
  \overset{(e)}{\approx} &   2\alpha \sqrt{{P}_l{P}_1(\Omega_\text{LO})}\Big[\cos[\phi_l - \phi_p(\Omega_\text{LO})] \\
    -&\kappa_2(\Omega_\text{LO})\cos\varphi_2(\Omega_\text{LO})U_x\cos(2\pi f_\delta t+\theta_\delta) \Big] \\
    +& \xi^{(B)}(t)\sqrt{G \alpha} \sqrt{{P}_l +{P}_1(\Omega_\text{LO})}\Big[1 -\frac{P_1(\Omega_\text{LO})}{P_l+P_1(\Omega_\text{LO})} \\
    \times &\kappa_1(\Omega_\text{LO})U_x\cos(2\pi f_{\delta}t+\theta_{\delta})\Big],
\end{split}
\end{equation}
where \textcolor{black}{\(\xi^{(B)}(t) = \frac{\xi_{+}(t)-\xi_{-}(t)}{\sqrt{2}}\sim \mathcal{N}(0,\varsigma^2)\) represent the differential combining, and (e) holds only when \(\Omega_{\text{LO}}\gg \Omega_{x}\)}. \(\kappa_2(\Omega_\text{LO})\). \(\varphi_{2}(\Omega_{\text{LO}})\), \(\varphi_{2}(\Omega_{\text{LO}})\), and \(\Re\{\chi^{\prime}(\Omega_{\text{LO}})\}\) are given by
\begin{equation}
    \kappa_{2}(\Omega_{\text{LO}})=\frac{\pi d\mu_{34}}{\lambda_{p}\hbar}\sqrt{\left[\Im\{\chi^{\prime}(\Omega_{\text{LO}})\}\right]^{2}+\left[\Re\{\chi^{\prime}(\Omega_{\text{LO}})\}\right]^{2}},
\end{equation}
\begin{equation}
    \varphi_{2}(\Omega_{\text{LO}})=\phi_{l}-\phi_{p}(\Omega_{\text{LO}})+\psi_{p}(\Omega_{\text{LO}}),
\end{equation}
\begin{equation}
    \psi_{p}(\Omega_{\text{LO}})=\arccos\frac{\Im\{\chi^{\prime}(\Omega_{\text{LO}})\}}{\sqrt{\left[\Im\{\chi^{\prime}(\Omega_{\text{LO}})\}\right]^{2}+\left[\Re\{\chi^{\prime}(\Omega_{\text{LO}})\}\right]^{2}}},
\end{equation}
and
\begin{equation}
    \begin{split}
        \Re\{\chi^{\prime}(\Omega_{\text{LO}})\}=-\frac{4N_{0}\mu_{12}^{2}}{\epsilon_{0}\hbar}\Omega_{\text{LO}}\left[\frac{2A_{1}\Omega_{\text{LO}}^{2}+A_{2}}{C_{1}\Omega_{\text{LO}}^{4}+C_{2}\Omega_{\text{LO}}^{2}+C_{3}}\right.\\
\left.-\frac{(A_{1}\Omega_{\text{LO}}^{4}+A_{2}\Omega_{\text{LO}}^{2}+A_{3})\left(2C_{1}\Omega_{\text{LO}}^{2}+C_{2}\right)}{(C_{1}\Omega_{\text{LO}}^{4}+C_{2}\Omega_{\text{LO}}^{2}+C_{3})^{2}}\right].
    \end{split}
\end{equation}

\subsection{Down-Conversion}
Based on the above results, we extract the time-varying signal from \( V^{(\mathrm{D})}(t)\) and \( V^{(\mathrm{B})}(t)\). Generally, the output voltage is given by
\begin{equation}
    \begin{split}
   \tilde V(t) &=2 \textcolor{black}{\alpha}\sqrt{G} {P}_G {\kappa}(\Omega_\text{LO})\cos\varphi(\Omega_\text{LO})U_x\cos(2\pi f_{\delta}t+\theta_{\delta}) \\
    & + \xi_\text{SN}(t) \sqrt{G\alpha} {\bar P}_\text{SN}\kappa_1(\Omega_\text{LO})U_x \cos(2\pi f_{\delta}t+\theta_{\delta}) \\
    &- \xi_\text{CN}(t)\sqrt{G \alpha{ \bar P}_\text{CN}},\\
\end{split}
\end{equation}
where we have \({P}_G \in \big\{P_1(\Omega_\text{LO}),  \sqrt{{P}_l{P}_1(\Omega_{\text{LO}})}\big\}\), 
\( {\kappa}(\Omega_\text{LO}) \in \big\{\kappa_1(\Omega_\text{LO}),\kappa_2(\Omega_\text{LO})\big\}\), \(\varphi(\Omega_\text{LO}) \in \big\{0,\varphi_2(\Omega_\text{LO}) \big\}\), \( \bar P_\text{CN} \in \big\{{P_{1}(\Omega_\text{LO})},{{P_l+P_1(\Omega_\text{LO})}} \big\}\), and \(\bar P_{\text{SN}} \in \big\{\sqrt{P_{1}(\Omega_\text{LO})},\frac{P_{1}(\Omega_\text{LO})}{\sqrt{P_l+P_1(\Omega_\text{LO})}} \big\}\) related to the DIOD and the BCOD, respectively. \(\xi_\text{SN}(t)\)  and \(\xi_\text{CN}(t)\) denote the baseband part and the narrowband component near \(f_\delta\), respectively. Besides, these two filtered components are independent \footnote{The same underlying Gaussian fluctuation process can be decomposed into two spectrally disjoint components. The low-frequency component is upconverted by the signal-dependent term and retained by the first alternating current-preserving filter, whereas the narrowband component centered at \(f_\delta\) passes directly through the same filter and constitutes the direct-current-dependent noise. Under the Gaussian assumption and the condition of non-overlapping passbands, these two filtered components are independent. Therefore, we model the photodetector noise as wideband white Gaussian.}. As can be seen, the output voltage consists of three parts, including the user's signal, user-signal-dependent noise, and direct-current-dependent noise. By passing through the homodyne receiver and a low-pass filter, we obtain the I/Q signal, which is given by
\begin{equation}
    \begin{split}
         \tilde V^I_b(t) &=\textcolor{black}{2\alpha} \sqrt{G} {P}_G {\kappa}(\Omega_\text{LO})\cos\varphi(\Omega_\text{LO})U_x\cos(\theta_{\delta})\\
         & + \xi_\text{SN}(t) \sqrt{G\alpha} {\bar P}_\text{SN}\kappa_1(\Omega_\text{LO})U_x \cos(\theta_{\delta})+ w_I(t),\\
    \end{split}
\end{equation}
\begin{equation}
    \begin{split}
          \tilde V^Q_b(t) &=\textcolor{black}{2\alpha}\sqrt{G} {P}_G  {\kappa}(\Omega_\text{LO})\cos\varphi(\Omega_\text{LO})U_x\sin(\theta_{\delta})\\
         & + \xi_\text{SN}(t)\sqrt{G\alpha} {\bar P}_\text{SN}\kappa_1(\Omega_\text{LO})U_x \sin(\theta_{\delta})+w_Q(t),\\
    \end{split}
\end{equation}
where $w_I(t)\triangleq {\rm LPF}\{\xi_\text{CN}(t)\sqrt{G \alpha{ \bar P}_\text{CN}}\cos(2\pi f_\delta t)\}$ and
$w_Q(t)\triangleq {\rm LPF}\{-\xi_\text{CN}(t)\sqrt{G \alpha{ \bar P}_\text{CN}}\sin(2\pi f_\delta t)\}$ are the equivalent complex baseband noises. Importantly, after down-conversion and LPF, the direct-current-dependent noise becomes baseband noise.

\subsection{Equivalent Complex model of Signal-Dependent Shot noise}
Defining $ \sqrt{2}\tilde V_b(t)\triangleq \tilde V^I_b(t)+j \tilde V^Q_b(t)$ and using  \(P_x = \frac{1}{2}c\epsilon_0A_e |U_x|^2\), the complex-baseband signal is 
\begin{equation}
    \begin{split}
         \tilde V_b(t)  & = \frac{2\sqrt{G}\alpha}{\sqrt{c\epsilon_0 A_e}}{P}_G  {\kappa}(\Omega_\text{LO})\cos\varphi(\Omega_\text{LO})e^{-j\theta_\text{LO}}\sqrt{P_x}e^{j\theta_x}  \\
    & + \xi_\text{SN}\frac{\sqrt{G\alpha}}{\sqrt{c\epsilon_0 A_e}}{\bar P}_{\text{SN}}\kappa_1(\Omega_\text{LO})e^{-j\theta_\text{LO}}\sqrt{P_x}e^{j\theta_x}+w(t) \\
    & \triangleq \sqrt{\frac{\rho}{A_e}}\Phi x_b(t) + \xi_\text{SN} \sqrt{\frac{\rho_\text{SN}}{A_e}}\Phi_\text{SN} x_b(t)+w(t)
    \end{split}
\end{equation}
where \(w(t) =w_I(t) + jw_Q(t) \) and \( \Phi_\text{SN}= e^{-j\theta_\text{LO}}\). \(\rho\), \(\Phi\), and \(\rho_\text{SN}\) are given by
\begin{equation}
    \begin{split}
        \rho = 4GZ_0\alpha^2\left\{\begin{array}{cc}
            P^2_1(\Omega_\text{LO})\kappa^2_1(\Omega_\text{LO}), & \text{DIOD}\\
            P_lP_1(\Omega_\text{LO})\kappa^2_2(\Omega_\text{LO}), & \text{BCOD}
        \end{array}\right.,
    \end{split}
\end{equation}
\begin{equation}
    \begin{split}
        \Phi= \left\{\begin{array}{cc}
           e^{-j\theta_\text{LO}}, & \text{DIOD}\\
             e^{-j\theta_\text{LO}}\cos\varphi_2(\Omega_\text{LO}), & \text{BCOD}
        \end{array}\right.,
    \end{split}
\end{equation}
\begin{equation}
    \begin{split}
         \rho_\text{SN} = GZ_0\alpha\left\{\begin{array}{cc}
            P_1(\Omega_\text{LO})\kappa^2_1(\Omega_\text{LO}), & \text{DIOD}\\
            \frac{P^2_1(\Omega_\text{LO})}{P_l+P_1(\Omega_\text{LO})}\kappa^2_1(\Omega_\text{LO}), & \text{BCOD}
        \end{array}\right.,
    \end{split}
\end{equation}

Let us assume that the received information signal is \(x_b(t) = \sqrt{A_e}\sum_{\ell} h_\ell(t) s_b(t-\tau_\ell)\), where \(s_b(t)\) is the transmitted information, \(\tau_\ell\) and \(  h_\ell(t)\) denote the delay and time-varying gain of the \(\ell\)-th path, respectively. Finally, we arrive at the equivalent SISO model with signal-dependent noise, which is given by
\begin{equation}
    \begin{split}
         \tilde{V}_b(t) &= \sqrt{\rho}\Phi\sum_\ell h_\ell(t)s_b(t-\tau_\ell) \\
        & + \xi_\text{SN}  \sqrt{\rho_{\text{SN}}} \Phi_{\text{SN}} \sum_\ell h_\ell(t)s_b(t-\tau_\ell) + w(t),
    \end{split}
\end{equation}
where \(w(t)\) denotes the sum of direct-current-dependent noise, thermal noise, and quantum projection noise (QPN). 

We then explain the noise model in detail.
Firstly, the resulting direct-current-dependent noise within the bandwidth $B$ and the average photocurrent \(\bar P_\text{CN}\) is 
\begin{equation}
  \textcolor{black}{N_\text{CN} = \varsigma^2G\alpha \bar P_\text{CN}}.
\end{equation}
It is worth emphasizing that raw photodetector shot noise is a Poisson point process, rather than inherently Gaussian or complex-valued. However, its circularly symmetric complex Gaussian (CSCG) baseband representation arises from the following three reasons:
\begin{enumerate}
    \item For large photon counts, the Poisson distribution converges to a Gaussian by the central limit theorem.
    \item After bandpass filtering around the intermediate frequency \(f_\delta\), after bandpass filtering, the resulting narrowband real-valued noise can be represented by statistically independent in-phase and quadrature components with equal variance. 
    \item Quadrature downconversion and low-pass filtering then yield a complex-valued noise process whose real and imaginary parts are independent and identically distributed zero-mean Gaussians, which is precisely the definition of CSCG. 
\end{enumerate}
Secondly, the thermal noise is related to the temperature \(T\) and bandwidth \(B\), denoted as \(N_\text{TN} = k_BTBG\). The QPN is modeled as \cite{gong2026rydbergatomicquantumreceivers}
\begin{equation}
  \textcolor{black}{N_\text{QPN} = \rho c\epsilon_0 \cos^2\varphi(\Omega_\text{LO})B\frac{\hbar^2}{NT_2\mu^2_{34}},}
\end{equation}
where \(T_2\) and  \(N\)  mean the coherence time of the EIT process and the quantum mechanically uncorrelated atoms, respectively. Finally, by the superposition of the direct-current-dependent shot noise, thermal noise, and the QPN, the aggregate equivalent noise can be modeled as a complex additive white Gaussian noise process with variance
\begin{equation}
N_{\rm sum}=\frac{N_{\rm CN}+N_{\rm QPN}+N_{\rm TN}}{2}.
\end{equation}

\subsection{Analysis for System Design}
The coupled optical dependencies in Eqs. (25)–(27) obscure the effect of individual design parameters on the received SNR. To extract closed-form insight, we first study the ideal resonance case, i.e, zero detuning \(\Delta_p = \Delta_c = \Delta_\text{RF} = 0\) , which serves as an analytical benchmark and admits tractable stationary-point analysis. Therefore, we have \(A_1 = A_2 = A_3 = B_2 = B_3= C_3 = 0\), \(B_1 =\gamma_2\), \(C_1 = 2\Omega^2_p + \gamma^2_2\), and \(C_2 = 2\Omega^2_c\Omega^2_p + 2\Omega^4_p \). Substituting these results into the previous expressions, we have
\begin{equation}
    \begin{split}
        \rho_{21}(\Omega_{\mathrm{RF}})&=\frac{-j \gamma_2\Omega_{p}\Omega_{\mathrm{RF}}^{2}}{(2\Omega^2_p + \gamma^2_2)\Omega_{\mathrm{RF}}^{2}+(2\Omega^2_c\Omega^2_p + 2\Omega^4_p)},\\
    \end{split}
\end{equation}
\begin{equation}
    \kappa_1(\Omega_\text{LO}) = \kappa_2(\Omega_\text{LO}) = \frac{\pi d \mu_{34}}{\lambda_p \hbar}\Im\{\chi^{\prime}(\Omega_{\text{LO}})\},
\end{equation}
\begin{equation}
    \begin{split}
        \Im\{\chi^{\prime}(\Omega_{\text{LO}})\}&=\frac{4N_{0}\mu_{12}^{2}}{\epsilon_{0}\hbar}\frac{\Omega_{\text{LO}}\gamma_2(2\Omega^2_c\Omega^2_p + 2\Omega^4_p)}{[(2\Omega^2_p + \gamma^2_2)\Omega_{\text{LO}}^{2}+(2\Omega^2_c\Omega^2_p + 2\Omega^4_p )]^2},
    \end{split}
\end{equation}
\begin{equation}
    \Re\{\chi^{\prime}(\Omega_{\text{LO}})\}=0, \psi_{p}(\Omega_{\text{LO}}) = 0,
\end{equation}
\begin{equation}
    \varphi_{2}(\Omega_{\text{LO}})=\phi_{l}-\phi_0.
\end{equation}

Then, by using \(\Omega_c = \frac{\mu_{23}}{\hbar}U_c\), \( \Omega_\text{LO} = \frac{\mu_{34}}{\hbar}U_{\text{LO}}\), \(\Omega_p = \frac{\mu_{12}}{\hbar}U_0\), \(P_c = \frac{\pi c \epsilon_0}{8 \ln 2 }F^2_c |U_c|^2\), \(P_0 = \frac{\pi c \epsilon_0}{8 \ln 2 }F^2_p |U_0|^2\), and \(P_{\text{LO}} = \frac{1}{2}c\epsilon_0 A_e |U_{\text{LO}}|^2\) and defining \(a_{12} \triangleq \frac{\mu_{12}^2}{\hbar^2}\frac{8\ln 2}{\pi c\epsilon_0 F^2_p} \), \(a_{23} \triangleq \frac{\mu_{23}^2}{\hbar^2}\frac{8\ln 2}{\pi c\epsilon_0 F^2_c}\), \(a_{34} \triangleq \frac{\mu_{34}^2}{\hbar^2}\frac{2}{c\epsilon_0 A_e}\), and \(C_0 \triangleq \frac{8\pi d N_0\mu_{12}^2\mu_{34}\gamma_2}{\epsilon_0\lambda_p\hbar^2}\), we have \
\begin{equation}
    \begin{split}
        \label{P_1Pb}
       P_1(\Omega_\text{LO}) = P_0  \exp \Bigg(
\frac{-4\pi d\,N_0 \mu_{12}^2 \gamma_2 a_{34}P_\text{LO}}
{\epsilon_0 \hbar \lambda_p X_1}
\Bigg),
    \end{split}
\end{equation}
\begin{equation}\label{Omega_1Pb}
 \kappa_1(\Omega_\text{LO}) = \kappa_2(\Omega_\text{LO}) = C_0 \frac{\sqrt{ a_{34}P_{\text{LO}}}a_{12}P_0\big(a_{23}P_c+a_{12}P_0\big)}{X^2_1},
    \end{equation}
where  \(X_1 = 2a^2_{12} P_0^2 + 2a_{12} P_0(a_{34}P_\text{LO}+a_{23}P_c) + \gamma_2^2a_{34}P_{\text{LO}}\). Furthermore, by judiciously designing the LO and probe beam, we can readily have \(\phi_l = \phi_0\), i.e., \(\Phi = \Phi_\text{SN} = e^{-j\theta_\text{LO}}\). 

In the following, the stationary-point-based design insights under ideal resonance conditions are provided. For ease of derivation, we denote the normalized noise as \(\mathcal{W} \triangleq  \frac{D_1\bar{P}^2_\text{SN} }{ {P}_G^2}
+ \frac{D_2\bar{P}_\text{CN}}{ {P}_G^2\kappa^2(\Omega_\text{LO})}+\frac{D_3}{{P}_G^2\kappa^2(\Omega_\text{LO})} +  D_4\), where \(D_1\), \(D_2\), \(D_3\), and \(D_4\) represent the constant coefficients. As a result, the optimal design can be obtained by setting the first-order derivative to 0, i.e., \(    \frac{\partial \mathcal{W}}{P_{\text{F}}} = 0\), \(\text{F} \in \{l, \text{LO}, c, 0\}\). 

\begin{lemma}
    \label{Lemoptical}
    With the given probe beam \(P_0\), coupling beam \(P_c\), and LO signal \(P_\text{LO}\), the optimum local optical beam power is
\( P_l^{\star} = \min\left\{\,P_l^{\max}, I_{\mathrm{sat}}/\alpha - P_1(\Omega_{\mathrm{LO}})\right\}\), where \(P_l^{\max}\) is the beam's maximum power and \(I_{\mathrm{sat}}\) is the photodetector saturation current.

     \emph{Proof}: The first-order partial derivative
 \(\frac{\partial \mathcal{W}}{P_l} =
-\,\frac{D_1P_1(\Omega_{\rm LO})\big(2P_l+P_1(\Omega_{\rm LO})\big)}
{P_l^2\big(P_l+P_1(\Omega_{\rm LO})\big)^2}
-\frac{D_2}{\kappa_1^2(\Omega_{\rm LO})P_l^2}
-\frac{D_3}{P_l^2P_1(\Omega_{\rm LO})\kappa_1^2(\Omega_{\rm LO})}\) is strictly negative for \(P_l > 0\), and thus \(\mathcal{W}\) is a decreasing function of \(P_l\). $\hfill\blacksquare$
\end{lemma}

However, \(P_1(\Omega_\text{LO})\) and \(\kappa(\Omega_\text{LO})\) highly depends on the power of the LO, coupling beam, and probe beam, and thus it is challenging to obtain the optimal design. Instead, we further discuss the relationship between the three noises and the beams (LO signal).
\subsubsection{User-Signal-Dependent Noise}
We characterize the impacts of system design on signal-dependent noise \(\mathcal{N}_1 =  \frac{D_1\bar{P}^2_{\text{SN}} }{ {P}_G^2}\). Firstly, taking the logarithm of \(P_1(\Omega_\text{LO})\), we have
\begin{equation}
\begin{split}
     \ln{P_1(\Omega_\text{LO})} = \ln P_0 -X_0 \frac{a_{34}P_\text{LO}}{X_1}, \\
\end{split}
\end{equation}
where \(X_0 =  \frac{4\pi d\,N_0 \mu_{12}^2 \gamma_2 }
{\epsilon_0 \hbar \lambda_p}\) is the constant term. For DIOD detection, we observe that minimizing the signal-dependent noise is equivalent to maximizing \(P_1(\Omega_\text{LO})\). Using \( f(x)\frac{\partial \ln f(x)}{\partial x} = \frac{\partial f(x)}{\partial x}\), the first-order partial derivative of \(\ln P_1(\Omega_\text{LO})\) is given by
\begin{equation}\label{sdpart}
    \begin{split}
        \frac{\partial \ln P_1(\Omega_\text{LO})}{\partial P_F} = \left\{\begin{array}{cc}
        -\frac{X_0a_{34}}{X^2_1} (2a^2_{12} P_0^2 +2a_{12} P_0a_{23}P_c),   & P_\text{LO} \\ 
        \frac{X_0a_{34}P_\text{LO}}{X^2_1}2a_{12}a_{23}P_0,  & P_c \\
       \frac{1}{P_0} + \frac{X_0a_{34}P_\text{LO}}{X^2_1}\frac{\partial X_1}{\partial P_0},    & P_0
        \end{array}\right.
    \end{split},
\end{equation}
where \(\frac{\partial X_1}{\partial P_0} = 4a^2_{12} P_0+2a_{12}(a_{34}P_\text{LO}+a_{23}P_c) \). Since \(P_0\), \(X_0\), \(X_1\), \(P_\text{LO}\), and \(P_c \) are larger than 0, the SNR under the user signal-dependent noise decreases with LO power but increases with coupling/probe power. Regarding the BCOD scheme, the first-order partial derivative of \(\ln(\mathcal{N}_1)\) is 
\begin{equation}\label{BCODSN}
\begin{split}
        \frac{\partial \ln(\mathcal{N}_1)}{\partial P_F} 
        & = \frac{P_lP_1(\Omega_\text{LO})}{P_1(\Omega_\text{LO}) + P_l}\frac{\partial \ln P_1(\Omega_\text{LO})}{\partial P_F}.
\end{split}
\end{equation}
As can be seen from \eqref{BCODSN}, the optimal design for BCOD schemes is opposite to that for DIOD detection.
As a result, the SNR under signal-dependent noise is minimized at the boundary \(\{P^{\max}_{\text{LO}}, P^{\min}_c, P^{\min}_0\}\) within a linear reception regime. 
\subsubsection{Direct-Current-Dependent Noise} Defining the second term as \({\mathcal{N}_2} = \frac{D_2\bar{P}_\text{CN}}{{P}_G^2\kappa^2(\Omega_\text{LO}) }\), minimizing noise caused by direct current can be obtained by maximizing \(\frac{1}{\mathcal{N}_2} = \frac{ {P}_G^2\kappa^2(\Omega_\text{LO}) }{D_2\bar{P}_\text{CN}}\). By using \(\ln(\frac{1}{\mathcal{N}_2}) = 2\ln{{P}_G} + 2\ln{\kappa(\Omega_\text{LO})}-\ln{\bar{P}_\text{CN}}\), the first-order partial derivative with respect to \(P_F\), \(\text{F} \in \{\text{LO},c,0\}\), is 
\begin{equation}
    \frac{\partial \ln(\frac{1}{\mathcal{N}_2}) }{\partial P_\text{F} } = 2 \frac{\partial\ln{{P}_G}}{\partial P_\text{F}} + 2 \frac{\partial\ln{\kappa(\Omega_\text{LO})}}{\partial P_\text{F}}-\frac{\partial\ln{\bar{P}_\text{CN}}}{\partial P_\text{F}}.
\end{equation}
Setting the partial derivative to 0, the optimal coupling is given by
\begin{equation}
\begin{split}\label{optimalcou}
        P^*_{c,\text{CN}} = \frac{a_{34}P_\text{LO}(\Gamma X_0 + \sqrt{\Gamma^2X_0^2+16X^2_2})}{8a_{12}a_{23}P_0}-\frac{a_{12}P_0}{a_{23}}   , 
\end{split}
\end{equation}
where \(X_2 = 2a_{12}P_0+ \gamma^2_2\) and \(\Gamma = 1\) for the DIOD scheme or \(\Gamma = \frac{P_l}{P_l+P_1(\Omega_\text{LO}) }\) for BCOD detection. The optimal LO signal is given by
 \begin{equation} \label{optiamlLO}
    P^*_{\text{LO},\text{CN}} =\frac{2a_{12}P_0(a_{23}P_c+a_{12}P_0)\sqrt{X_3}-\Gamma X_0 - 2X_2}{6a_{34}X^2_2},
\end{equation}
where \(X_3 = \Gamma^2X_0^2 + 4\Gamma X_0X_2+16X^2_2\). \textcolor{black}{Note that \eqref{optimalcou} and \eqref{optiamlLO} are exact closed-form solutions for DIOD, but not for BCOD. Nevertheless, under the typical weak-probe condition \(P_l \gg P_1(\Omega_\text{LO})\), the BCOD expressions also admit an accurate approximate closed-form solution.}
The optimum design of the probe beam is challenging to solve, as the probe-beam optimization couples polynomial and exponential dependencies. To tackle this issue, we employ Newton’s method by setting \(\frac{\partial \mathcal{W}}{P_0} = 0\), which provides an efficient and numerically stable solution.

\subsubsection{Thermal Noise} Defining the normalized thermal noise as \({\mathcal{N}_3} = \frac{D_3}{{P}_G^2\kappa^2(\Omega_\text{LO}) }\), the system can be designed by maximizing \(\frac{1}{\mathcal{N}_3} = \frac{ {P}_G^2\kappa^2(\Omega_\text{LO}) }{D_3}\). The optimal coupling beam and LO signal for the DIOD detection are given by
\begin{equation}
    P^*_{c,\text{TN}} = \frac{a_{34}P_\text{LO} (X_0 + \sqrt{X_0^2 + 4X^2_2})}{4a_{12}a_{23}P_0} - \frac{a_{12}P_0}{a_{23}},
\end{equation}
and 
\begin{equation}
    P^*_{\text{LO},\text{TN}} = \frac{2a_{12}P_0(a_{23}P_c + a_{12}P_0) X_3}{3a_{34}X^2_2},
\end{equation}
where \(X_3 = \sqrt{(X_0+X_2)^2 + 3X^2_2}-X_0-X_2\). Furthermore, the closed-form expression of the probe beam can be obtained by Newton's method. For BCOD, we observe that the first-order partial derivative with respect to LO and coupling beam aligns with the direct-current-dependent noise scenario, and thus the optimum expressions of LO and coupling beam are given by \eqref{optimalcou} and \eqref{optiamlLO}, respectively.

Building upon the analysis, we can strike a good balance between the effective reception gain and the noise floor by judiciously designing the optical parameters. Consequently, a \emph{noise-composition-driven} beam/LO design guideline is summarized as follows:
\begin{enumerate}
\item \textbf{User-signal-dependent-noise regime}:
Since minimizing $\mathcal N_1$ is equivalent to maximizing $P_1(\Omega_\text{LO})$ and $P_1(\Omega_\text{LO})$ is monotonically decreasing with $P_{\rm LO}$ but increasing with $(P_c,P_0)$, the optimal design is attained at the feasible boundary subjecting to a linear-operation regime.

\item \textbf{Direct-current shot noise/thermal-noise  regime}:
In this regime, the design aims to maximize ${P}_G^2\kappa^2(\Omega_\text{LO})/\bar P_\text{CN}$ or ${P}_G^2\kappa^2(\Omega_\text{LO})$  by using the closed-form stationary points and solving $P_0$.

\item \textbf{Mixed-noise (balanced) regime}:
When $\mathcal N_1$ is comparable to shot/thermal noises, the optimal design is governed by the trade-off design between $P_1(\Omega_\text{LO})$ and $\kappa(\Omega_\text{LO})$.
Practically, this implies:
(i) $P_{\rm LO}^\star$ should lie between the minimum feasible LO and the shot/thermal stationary LO, i.e.,
$P_{\rm LO}^{\min}\le P_{\rm LO}^\star \lesssim P_{{\rm LO},{\rm TN}}^\star$;
(ii) $P_c^\star$ should be tuned around the shot/thermal stationary point to avoid the saturated region;
and (iii) $P_0$ is used to balance the noise composition.
\end{enumerate}

\section{Model of RAQ-MIMO with Signal-Dependent Noise}
In this section, we extend the SISO model to the general MIMO case and evaluate the impacts of signal-dependent noise on system performance.
\subsection{Extended to MIMO Model}
Building on the above, we generalize the SISO model to the MIMO system with an \(M\)-sensor BS and \(K\) single-antenna users. Similar to the previous derivations, the superimposed signal of \(K\) users at \(m\)-th vapor cell is given by
\begin{equation}
    \begin{split}
        x_{m}(t) = \sum\limits_{k=1}^K x_{m,k}(t) = \sum\limits_{k=1}^K{U_{x,m,k}}\cos(2\pi f_c t+\theta_{x,m,k}),
    \end{split}
\end{equation}
where \(U_{x,m,k}\), \(f_c\) and \(\theta_{x,m,k}\) are the electrical field, frequency, and phase shift, respectively.  
Similarly, the LO signal at the \(m\)-th vapor cell is expressed as 
\begin{equation}
    y_{m}(t) = {U_{\text{LO},m}}\cos(2\pi f_{\text{LO},m} t + \theta_{l,m}),
\end{equation}
where  \(U_{\text{LO},m}\), \(f_{\text{LO},m}\), and  \(\theta_{\text{LO},m}\) are the electrical field, the frequency, and the phase shift of the LO, respectively.
Then, the superimposed signal received at \(m\)-th vapor cell can be expressed as
\begin{equation}
\begin{split}
        z_m(t) &= y_{m}(t) + x_{m}(t) \approx {U_{z,m}}\cos(2 \pi f_{\text{LO},m} t+\theta_{\text{LO},m}),
\end{split}
\end{equation}
where \(U_{z,m}\) is the electrical field. According to \cite{gong2025rydbergatomicquantummimo}, \(U_{z,m}\) is given by
\begin{equation}
    \begin{split}
        U_{z,m} &=\sqrt{U_{\text{LO},m}^2 + 2U_{\text{LO},m}\sum\limits_{k=1}^KU_{x,m,k}\cos(2\pi f_{\delta,m}t + \theta_{\delta,m,k})}\\
        & \approx U_{\text{LO},m} + \sum\limits_{k=1}^KU_{x,m,k}\cos(2\pi f_{\delta,m}t + \theta_{\delta,m,k}),
    \end{split}
\end{equation}
where \(f_{\delta,m} = f_c - f_{\text{LO},m}\) and \(\theta_{\delta,m,k} = \theta_{x,m,k}-\theta_{\text{LO},m}\) mean the frequency and phase shift difference, respectively.

Then, by using the Rabi frequency and amplitude, the \(m\)-th sensor's Rabi frequency is given by
\begin{equation}
    \Omega_{\text{RF},m} \approx \Omega_{\text{LO},m} + \sum\limits\Omega_{x,m,k}\cos(2\pi f_{\delta,m} +\theta_{\delta,m,k}).
\end{equation}
By adopting a similar signal processing of SISO, the signal received at the \(m\)-th sensor is given by
\begin{equation}
    \begin{split}
         \tilde{V}_{m}(t) &= \sqrt{\rho_m}\Phi_m\sum\limits_{k=1}^K x_{m,k}(t) + \xi_{\text{SN},m}  \sqrt{\rho_{\text{SN},m}} \Phi_{\text{SN},m} \\
         &\times \sum\limits_{k=1}^K x_{m,k}(t)-  \xi_{\text{CN},m}\sqrt{G \alpha \bar P_{\text{CN},m}},
    \end{split}
\end{equation}
where \(\rho_m\) and \({\rho_{\text{SN},m}} \) denote the corresponding gain of \(m\)-th vapor cell. \(\Phi_m\) and \(\Phi_{\text{SN},m}\) are the phase shifts. \( \bar P_{\text{CN},m}\) is the power of the direct current at the  \(m\)-th sensor. Furthermore, we have theoretically identical gain \(\rho \triangleq \rho_m\) (\({\rho_{\text{SN}}}  \triangleq \rho_{\text{SN},m}\)) and power \( \bar P_{\text{CN}} \triangleq \bar P_{\text{CN},m} \), \(\forall m \in \{1,\cdots,M\}\). We assume the signal-dependent noise components are independent and identically distributed, i.e., \(\xi_{\text{SN},m} \sim \mathcal{N}(0,\varsigma^2)  \), \(\forall m \in \{1,\cdots, M\}\). \textcolor{black}{Additionally, the LO is supplied as a free-space RF plane wave radiated by a dedicated horn antenna placed in the far field of the sensor array, and thus it impinges on \(M\) Rydberg cells from a fixed angle of arrival \(\vartheta\). Consequently, the phase shift of the LO signal at \(m\)-th sensor is \(\theta_{l,m}=\theta_{l,1}+\frac{2\pi}{\lambda_{\text{LO}}}(m-1)d \sin\vartheta\), and thus \(\Phi_m = \Phi e^{-j\frac{2\pi}{\lambda_\text{LO}}(m-1)d \sin\vartheta}\) and \(\Phi_{\text{SN},m} = \Phi_{\text{SN}} e^{-j\frac{2\pi}{\lambda_\text{LO}}(m-1)d \sin\vartheta}\).}

\textcolor{black}{For ease of derivation, we collect the current-dependent shot noise, thermal noise, and quantum projection noise into the vector \(\mathbf{w}\).} The equivalent baseband channel for RAQ-MIMO with signal-dependent noise can be expressed as
\begin{equation}
    \mathbf{y}_e = \sqrt{\rho}\Phi \mathbf{D}\mathbf{H}\mathbf{P}\mathbf{s} + \sqrt{\rho_{\text{SN}}}\Phi_{\text{SN}}\mathbf{B}\mathbf{D}\mathbf{H}\mathbf{P}\mathbf{s} + \mathbf{w},
\end{equation}
where \(\mathbf{s} \in \mathbb{C}^{K\times 1}\) implies transmitted information, \(\mathbf{P} = \text{diag}\{\sqrt{p_1},\cdots,\sqrt{p_K}\}\) is transmission power, \(\mathbf{B} = \text{diag}\{\xi_{\text{SN},1} ,\cdots,\xi_{\text{SN},M}\}\) collects the factor of signal-dependent noise, and \(\mathbf{D} = \text{diag}\{1, e^{-j\frac{2\pi d}{\lambda}}\sin\vartheta,\cdots,e^{-j\frac{2\pi(M-1) d}{\lambda}}\sin\vartheta\}\) is the phase shift related to wavelength \(\lambda\) and distance \(d\) between each vapor cell. \(\mathbf{w} \in \mathbb{C}^{M \times 1}\) contains the current-dependent shot noise, the thermal noise, and quantum projection noise, which can be treated as a complex additive white Gaussian noise with variance of  \(\sigma^2 =  N_\text{sum} \) \cite{gong2025rydbergatomicquantummimo}.  \(\mathbf{H} = [\mathbf{h}_1,\cdots, \mathbf{h}_K] \in \mathbb{C}^{M \times K}\) denotes the effective channel, where the \(k\)-th user's channel vector \(\mathbf{h}_k\) follows the complex Gaussian distribution of \(\mathcal{CN}(\mathbf{0},\beta_k \mathbf{I}_M)\) and \(\beta_k\) is the large-scale fading factor.

\begin{figure}[t]
    \centering
    \begin{minipage}[t]{0.8\linewidth}
        \centering
        \includegraphics[width=\linewidth]{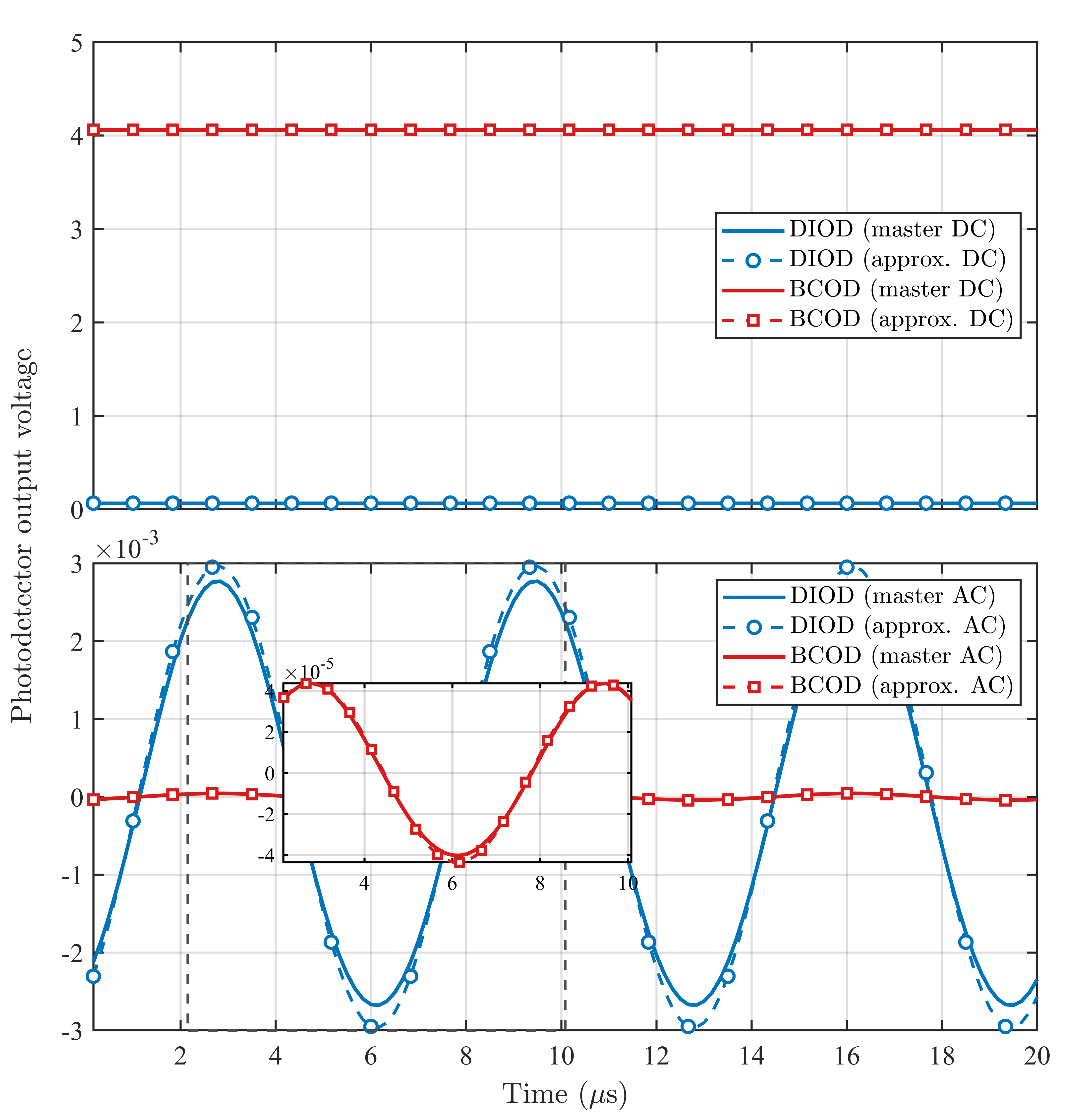}
        \caption{\textcolor{black}{Output voltage based on the master equation and approximated.}}
        \label{Waveform}
    \end{minipage}
\end{figure}

\begin{figure}[t]
    \centering
    \begin{minipage}[t]{0.8\linewidth}
        \centering
        \includegraphics[width=\linewidth]{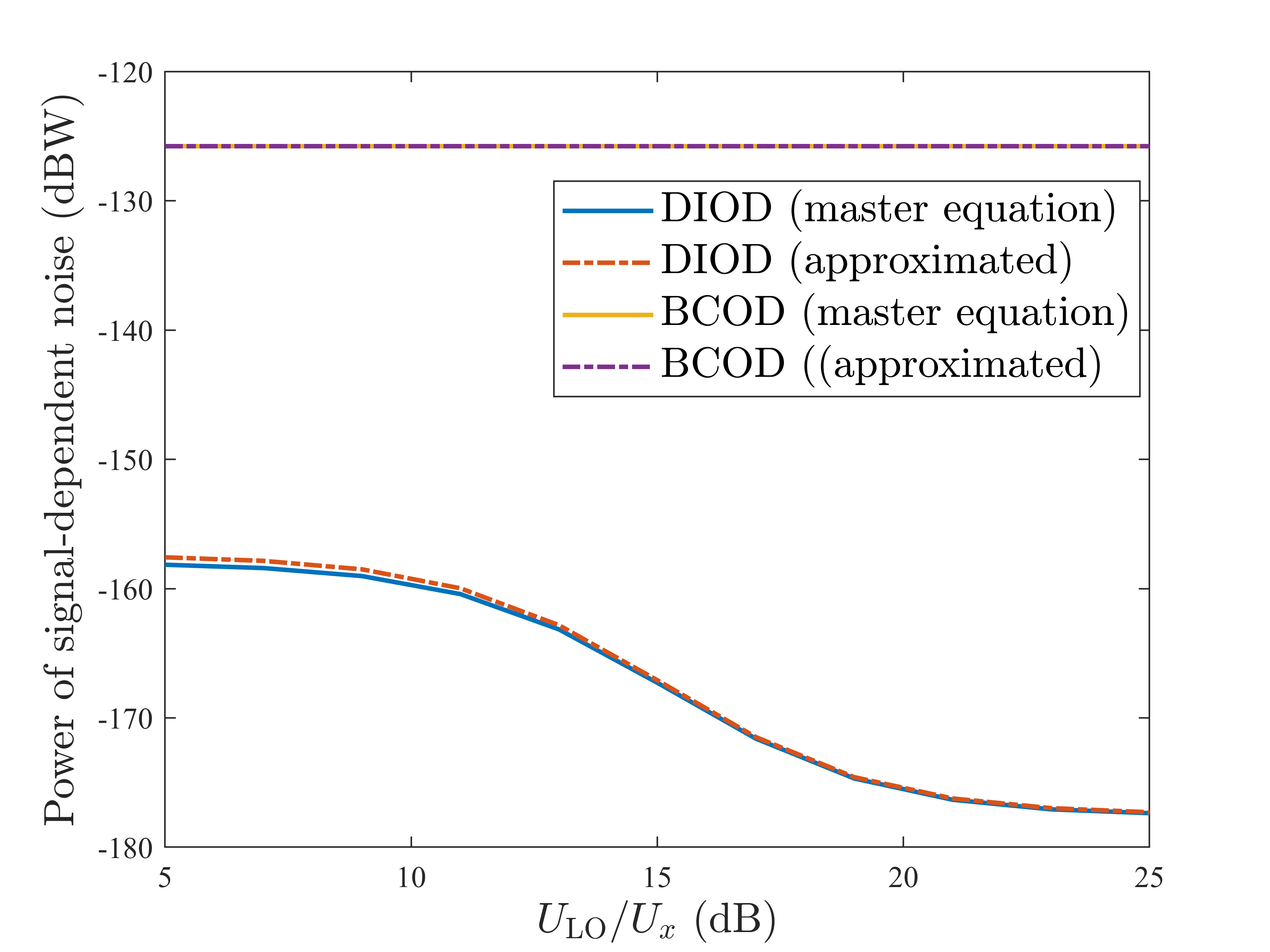}
        \caption{Signal-dependent shot noise with various powers of LO.}
        \label{powerSN}
    \end{minipage} 
\end{figure}

\subsection{Performance Analysis}
By assuming perfect channel state information, we evaluate the performance based on MRC and ZF detectors and denote \(\mathbf{C}^{\text{MRC}} = \Phi\mathbf{D}\mathbf{H}\) for MRC and \(\mathbf{C}^{\text{ZF}} = \Phi\mathbf{D}\mathbf{ H} [(\Phi\mathbf{D}\mathbf{ H})^H\Phi\mathbf{D}\mathbf{H}]^{-1}\) for ZF. \textcolor{black}{By utilizing two schemes and the use-and-then-forget (UatF) bounding technique \cite{marzetta2016fundamentals}, the received signal of the user \(k\) can be written as the sum of an effective desired signal, a leaked self-signal, multi-user interference, signal-dependent shot noise, and additive noise, which is given by}
\begin{equation}
    \label{rxsignal}
    \small
    \begin{split}
         r_k& = \underbrace{ \mathbb{E}\{\sqrt{\rho p_k}\Phi\mathbf{c}^{H}_k\mathbf{D}\mathbf{h}_k\}s_k}_{\text{Ds}_k} \\
        &+ \underbrace{ \sqrt{\rho p_k}\Phi\mathbf{c}^{H}_k\mathbf{D}\mathbf{h}_ks_k -\mathbb{E}\{\sqrt{\rho p_k}\Phi\mathbf{c}^{H}_k\mathbf{D}\mathbf{h}_k\}s_k}_{\text{Ls}_k}  \\
        & + \underbrace{\sum\limits_{k' \neq k}^K\sqrt{\rho p_{k'}}\Phi\mathbf{c}^{H}_k\mathbf{D}\mathbf{h}_{k'}s_{k'}}_{\text{UI}_{k,k'}}\\
        &+\underbrace{\sum\limits_{k'=1}^K \sqrt{\rho_{\text{SN}} p_{k'}}\Phi_{\text{SN}}\mathbf{c}^{H}_k\mathbf{B}\mathbf{D}\mathbf{h}_{k'}s_{k'} }_{\text{SN}_{k,k'}}+\underbrace{\mathbf{c}^{H}_k\mathbf{w} }_{\text{N}_k},
    \end{split}
\end{equation}
where \(\mathbf{c}_k\) is the \(k\)-th column of the detector \(\mathbf{C}^{\text{MRC}} \) or \(\mathbf{C}^{\text{ZF}} \), \(s_k \sim \mathcal{CN}(0,1)\) is the mutually independent data symbol, \(p_k\), \(\text{Ds}_k\), \(\text{Ls}_k\), \(\text{UI}_{k,k'}\), \(\text{SN}_{k,k'}\), and \(\text{N}_k\) represent the \(k\)-th user's transmit power, effective signal, leaked signal, inter-user interference, shot noise related to signal, and equivalent noise (including thermal noise, quantum projection noise, and shot noise related to beam), respectively. Accordingly, the achievable data rate \(R_k\) is \(R_k = \log_2(1+\text{SINR}_k)\), where \(\text{SINR}_k\) is expressed as
\begin{equation}
    \label{achRate}
    \begin{split}
            \text{SINR}_k =\frac{|\text{Ds}_k|^2}{|\text{Ls}_k|^2 + \sum\limits_{k' \neq k}^K|\text{UI}_{k,k'}|^2 + \sum\limits_{k' = 1}^K|\text{SN}_{k,k'}|^2 + |\text{N}_k|^2}.
    \end{split}
\end{equation}
As can be seen from \eqref{achRate}, we extend the UatF framework of  \cite{marzetta2016fundamentals} by introducing the additional term \(\text{SN}_{k,k'}\), which scales with the user transmit powers and cannot be absorbed into the constant additive noise. Furthermore, it is challenging to derive the closed-form expression of data rate. To tackle this issue, a tight analytical lower bound for the achievable rate of the user \(k\) is derived in the following theorems.

\begin{theorem}
\label{MRC_SINR_T}
The ergodic achievable rate for the $k$-th user employing the MRC decoder is lower bounded as
\begin{equation}
\setlength\abovedisplayskip{5pt}
\setlength\belowdisplayskip{5pt}
\small
\label{MRC_LB_rate}
R_k \ge {\underline R}_k^{\rm MRC} = \log_2(1+\text{SINR}^{\text{MRC}}_k),
\end{equation}
where \(\text{SINR}^{\text{MRC}}_k\) is given by \eqref{SINR_MRC}, which is shown at the bottom of the next page.

\begin{figure*}[hb]
\centering 
\hrulefill 
    \begin{equation}
\label{SINR_MRC}
\begin{split}
    &\text{SINR}^{\text{MRC}}_k =\frac{M\rho p_k |\Phi|^2\beta_k}{\sum\limits_{k'=1}^K p_{k'}\beta_{k'}(\rho|\Phi|^2+ \varsigma^2\rho_{\text{SN}}|\Phi_{\text{SN}}|^2) +  \varsigma^2\rho_{\text{SN}}p_k|\Phi_{\text{SN}}|^2\beta_k +\sigma^2}.
\end{split}
\end{equation}
\end{figure*}

\emph{Proof}: Please refer to Appendix \ref{Prooftheorem1}. $\hfill\blacksquare$
\end{theorem}

\begin{theorem}
\label{ZF_SINR_T}
For \(M > K\), the ergodic achievable rate for the \(k\)-th user employing the ZF decoder is lower bounded as
\begin{equation}
\setlength\abovedisplayskip{5pt}
\setlength\belowdisplayskip{5pt}
\small
\label{ZF_LB_rate}
R_k \ge {\underline R}_k^{\rm ZF} = \log_2(1+\text{SINR}^{\text{ZF}}_k),
\end{equation}
where \(\text{SINR}^{\text{ZF}}_k\) is given by
\begin{equation}
   \label{ZF_up_SINR}
   \begin{split}
          &\text{SINR}^{\text{ZF}}_k \\
          =& \frac{4(M-K)\rho p_k|\Phi|^2\beta_k}{\frac{\varsigma^2\rho_{\text{SN}}|\Phi_{\text{SN}}|^2}{M}\Big[p_k\beta_k(M-K) + \sum\limits_{k' = 1}^K p_{k'} \beta_{k'}(M-1)\Big] + \sigma^2}.
   \end{split}
\end{equation}

\emph{Proof}: Please refer to Appendix \ref{Prooftheorem2}. $\hfill\blacksquare$
\end{theorem}

Substituting the optical parameters into the derived lower bounds, we have \eqref{eq:SINR_MRC_D_expanded} and \eqref{eq:SINR_ZF_D_expanded}, which are given at the bottom of the next page. Items 1–5 correspond to inter-user interference, user-signal-dependent noise, direct-current-dependent noise, thermal noise, and quantum projection noise, respectively.
\begin{figure*}[hb]
\centering 
\hrulefill 
\begin{equation}\label{eq:SINR_MRC_D_expanded}
\mathrm{SINR}^{\mathrm{MRC}}_{k}
= \frac{
4Mp_k\beta_k
}
{\underbrace{4\sum\limits_{k'=1}^Kp_{k'}\beta_{k'}}_{\text{First term}} 
+\underbrace{\frac{\varsigma^2\bar{P}^2_{\text{SN}} }{\alpha {P}_G^2}\left(\sum\limits_{k'=1}^Kp_{k'}\beta_{k'} +p_k\beta_k\right)}_{\text{Second term}}
+ \underbrace{\frac{\varsigma^2 \bar{P}_\text{CN}}{2Z_0\alpha {P}_G^2\kappa^2(\Omega_\text{LO})}}_{\text{Third term}}+ \underbrace{\frac{k_B T B}{2Z_0\alpha^2  {P}_G^2\kappa^2(\Omega_\text{LO})}}_{\text{Fourth term}} +  \underbrace{\frac{2c\epsilon_0B\hbar^2}{NT_2\mu^2_{34}} }_{\text{Fith term}}}.
\end{equation}

\hrulefill 
  \begin{equation}\label{eq:SINR_ZF_D_expanded}
\mathrm{SINR}^{\mathrm{ZF}}_{k}
=
\frac{
4(M-K)p_k\beta_k }
{\underbrace{\Big[p_k\beta_k(M-K) + \sum\limits_{k' = 1}^K p_{k'} \beta_{k'}(M-1)\Big] \frac{\varsigma^2 \bar{P}^2_{\text{SN}}}{\alpha M {P}_G^2}}_{\text{Second term}}
+ \underbrace{\frac{\varsigma^2 \bar{P}_\text{CN}}{2Z_0\alpha {P}_G^2\kappa^2(\Omega_\text{LO})}}_{\text{Third term}}+\underbrace{\frac{k_B T B}{2Z_0\alpha^2 {P}_G^2\kappa^2(\Omega_\text{LO})}}_{\text{Fourth term}} +  \underbrace{\frac{2c\epsilon_0B\hbar^2}{NT_2\mu^2_{34}}}_{\text{Fifth term}}
}.
\end{equation}
\end{figure*}

\begin{remark}
    \emph{For the given system design, we observe that the power-scaling law still holds for the RAQ-MIMO systems. The non-zero rates of MRC and ZF are}
    \begin{equation}\label{asymptotic_rate}
     \log_2 \left(1+  \frac{4\beta_k}{\frac{\varsigma^2 \bar{P}_\text{CN}}{2Z_0\alpha {P}_G^2\kappa^2(\Omega_\text{LO})}+\frac{k_B T B}{2Z_0\alpha^2 {P}_G^2\kappa^2(\Omega_\text{LO})} +  \frac{2c\epsilon_0B\hbar^2}{NT_2\mu^2_{34}} }\right).
    \end{equation}
\end{remark}

\textcolor{black}{Although \eqref{asymptotic_rate} has the same form as the asymptotic rate of conventional RF MIMO, the mechanism in RAQ-MIMO is fundamentally different. The standard channel-hardening effect of MRC still suppresses inter-user interference at the rate \(\frac{1}{M}\). In contrast, the user-signal-dependent shot noise, which is unique to atomic reception, scales linearly with \(p_k\), and therefore vanishes as \(p_k \rightarrow 0\). Consequently, only the user-signal-independent noise floors, including direct-current shot noise, thermal noise, and the standard quantum limit, remain in \eqref{asymptotic_rate}. The asymptotic rate is thus governed primarily by the optical front end rather than the propagation channel.}

\begin{remark}
    \emph{Similar to the SISO case, the roles of the probe beam, coupling beam, and LO signal in RAQ-MIMO are not simply to increase the effective gain but to switch the operating point among the user-signal-dependent-noise-, direct-current-dependent-noise-, and thermal-noise-dominant regimes. Equivalently, RAQ-MIMO outperforms conventional RF-MIMO only when the optical operating point keeps the normalized user-signal-independent noise floor below the RF-MIMO thermal-noise floor, i.e.,} 
    \begin{equation}
        \frac{\varsigma^2\bar{P}_{\mathrm{CN}}}{2Z_0\alpha P_G^2\kappa^2(\Omega_{\mathrm{LO}})} + \frac{k_BTB}{2Z_0\alpha^2 P_G^2\kappa^2(\Omega_{\mathrm{LO}})} + \frac{2c\epsilon_0B\hbar^2 }{NT_2\mu_{34}^2}\;<\; 4\sigma_{\mathrm{RF}}^2.
    \end{equation}
\end{remark}

In particular, once the system is pushed into an unfavorable operating region, i.e., with an excessively large LO power, or an imbalanced probe-beam setting, the direct-current-dependent and thermal-noise terms may dominate, while the gain improvement becomes marginal due to the intrinsic gain-penalty trade-off.

\section{Simulation Results}
\begin{figure*}[ht]
 \centering
   \includegraphics[width=0.9\linewidth]{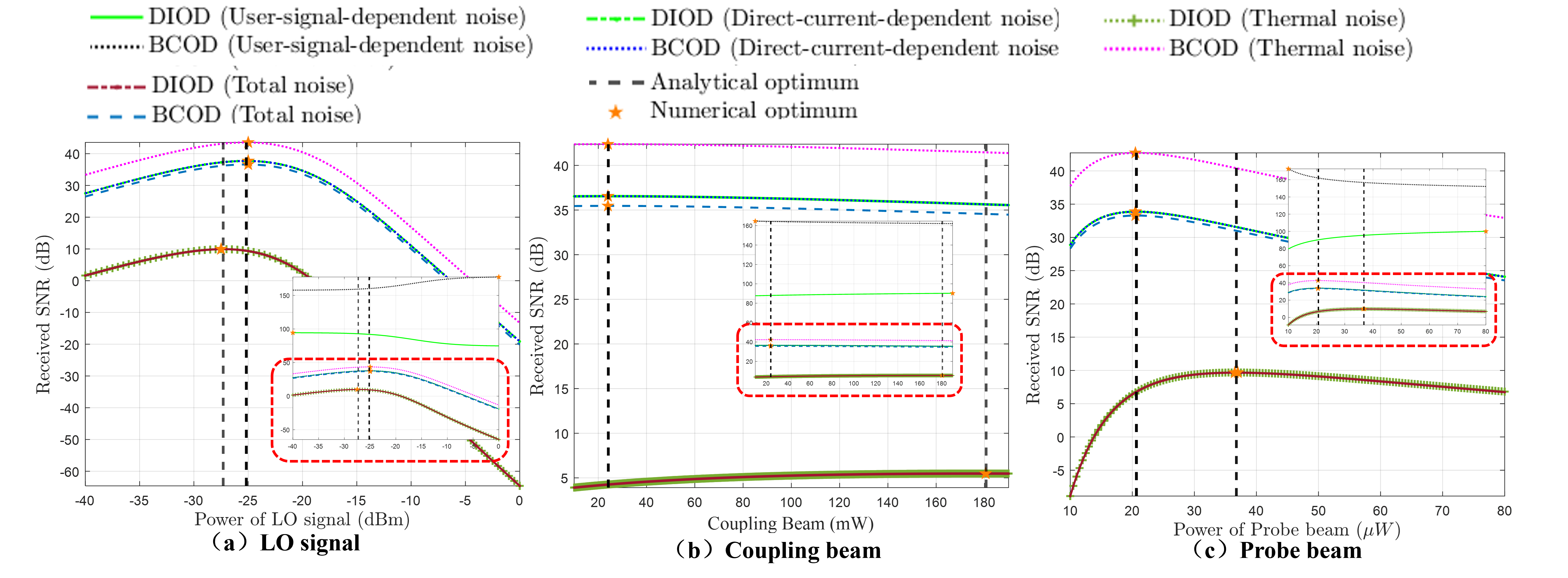}
       \caption{\textcolor{black}{The optimum design of the SISO-RAQR system under diverse noise sources (optima agree to within 0.1 dB).}}
    \label{SISOTN}
\end{figure*}

\begin{figure*}[ht]
   \centering
   \includegraphics[width=0.9\linewidth]{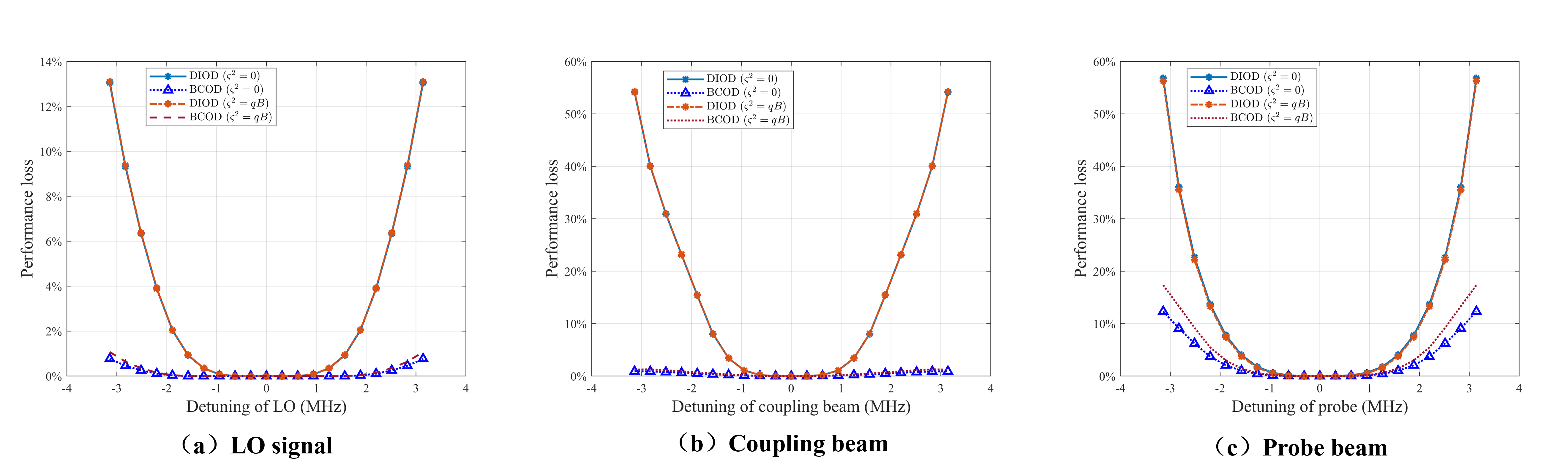}
    \caption{\textcolor{black}{The performance loss caused by detuning when the system is designed under perfect resonance (EIT linewidth \(\frac{\Omega^2_c}{\gamma_2}\approx 1.14\) MHz).}}
       
    \label{SISOloss}
\end{figure*}
In this section, we validate our approximated system model by simulations and then investigate the impact of the coupling/probe beam and LO signal on performance through simulation, thereby enabling a thorough analysis of the system's trade-off design.

\subsection{Parameter Settings}

Consider the four-level electron transition scheme of \(6{\rm S}_{1/2}\rightarrow6{\rm P}_{3/2}\rightarrow47{\rm D}_{5/2}\rightarrow48{\rm P}_{3/2}\) \textcolor{black}{with corresponding carrier frequency \(f_c = 6.9458\) GHz and narrow bandwidth \(B = 150\) kHz}. Unless otherwise stated, the quantum parameters are given in Table \(\textcolor{red}{\rm I}\) in \cite{gong2026rydbergatomicquantumreceivers}. It is assumed that \(K = 10\) devices are randomly distributed within a circular region of radius 50 meters, while their signals are collected by a BS with RAQRs located 1500 meters from the center of the device region. The large-scale fading factors (dB) can be obtained by using \(-32.4-20\lg(d_k)-20\lg(f_c)\), where \(d_k\) (in meters) is the distance between the device \(k\) and RAQR and \(f_c\) (in GHz) is the carrier frequency.
Furthermore, the simulation results are averaged over \(10^{4}\) realizations.

\subsection{SISO}

\begin{figure*}[ht]
    \centering
    \begin{minipage}[t]{0.42\linewidth}
        \centering
        \includegraphics[width=\linewidth]{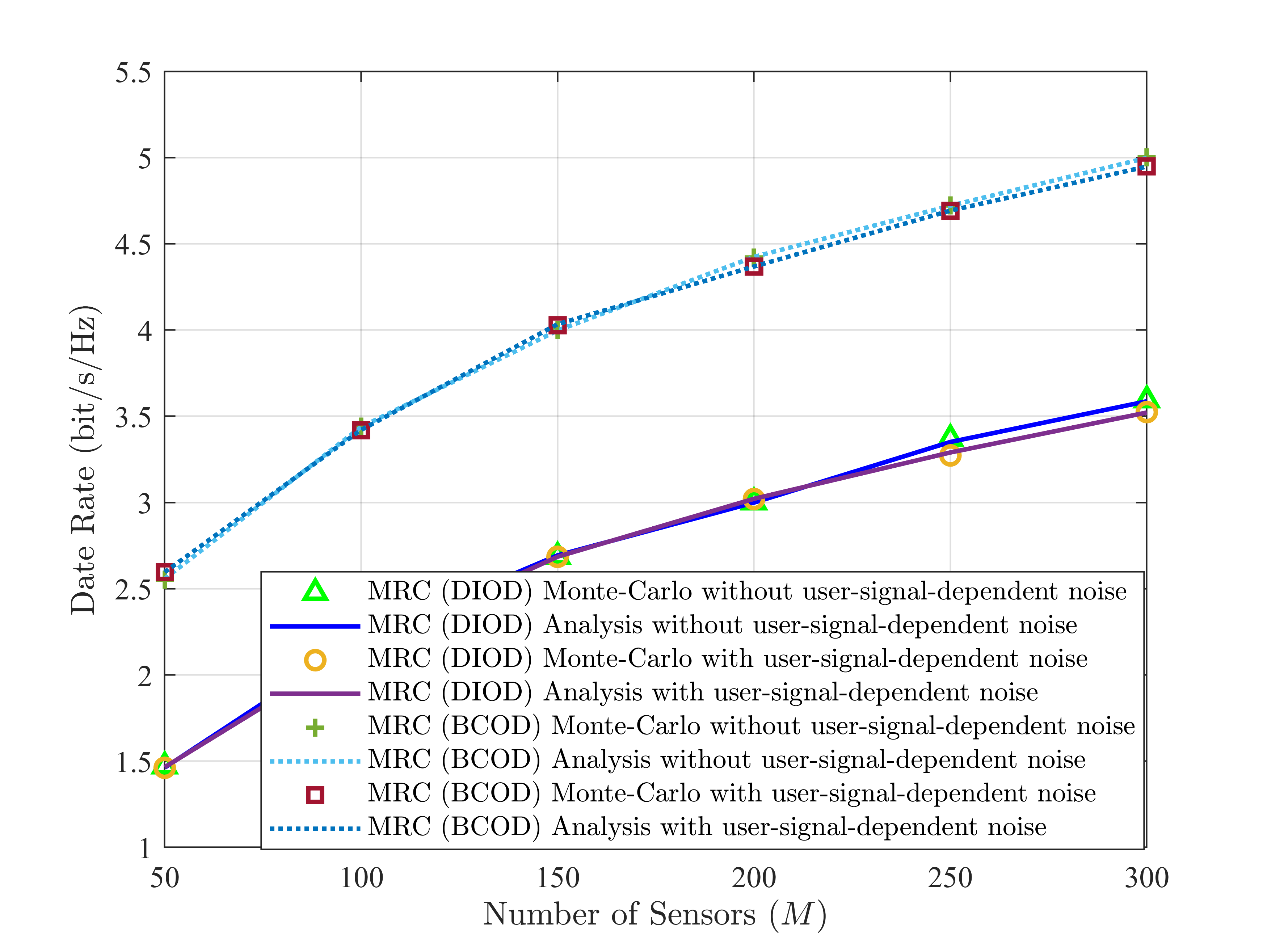}
        \subcaption{MRC detection.}
        \label{MRCvsM}
    \end{minipage}\hfill
    \begin{minipage}[t]{0.42\linewidth}
        \centering
        \includegraphics[width=\linewidth]{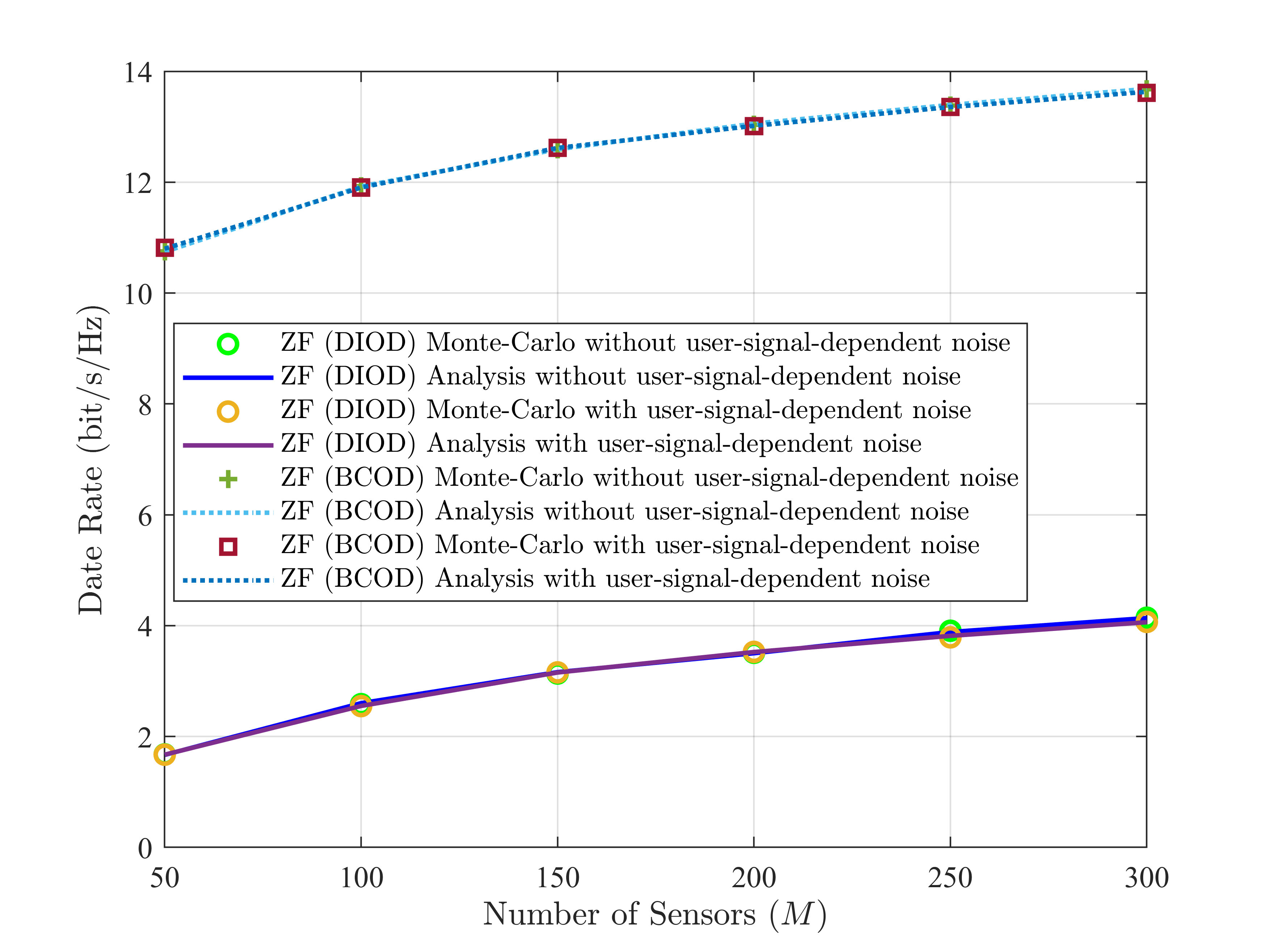}
         \subcaption{ZF detection.}
        \label{ZF}
    \end{minipage}\hfill

    \caption{Data rate of RAQ-MIMO under various atomic sensors \(M\).}
    \label{ratevsM}
\end{figure*}

\begin{figure*}[ht]
    \centering
    \begin{minipage}[t]{0.45\linewidth}
        \centering
        \includegraphics[width=\linewidth]{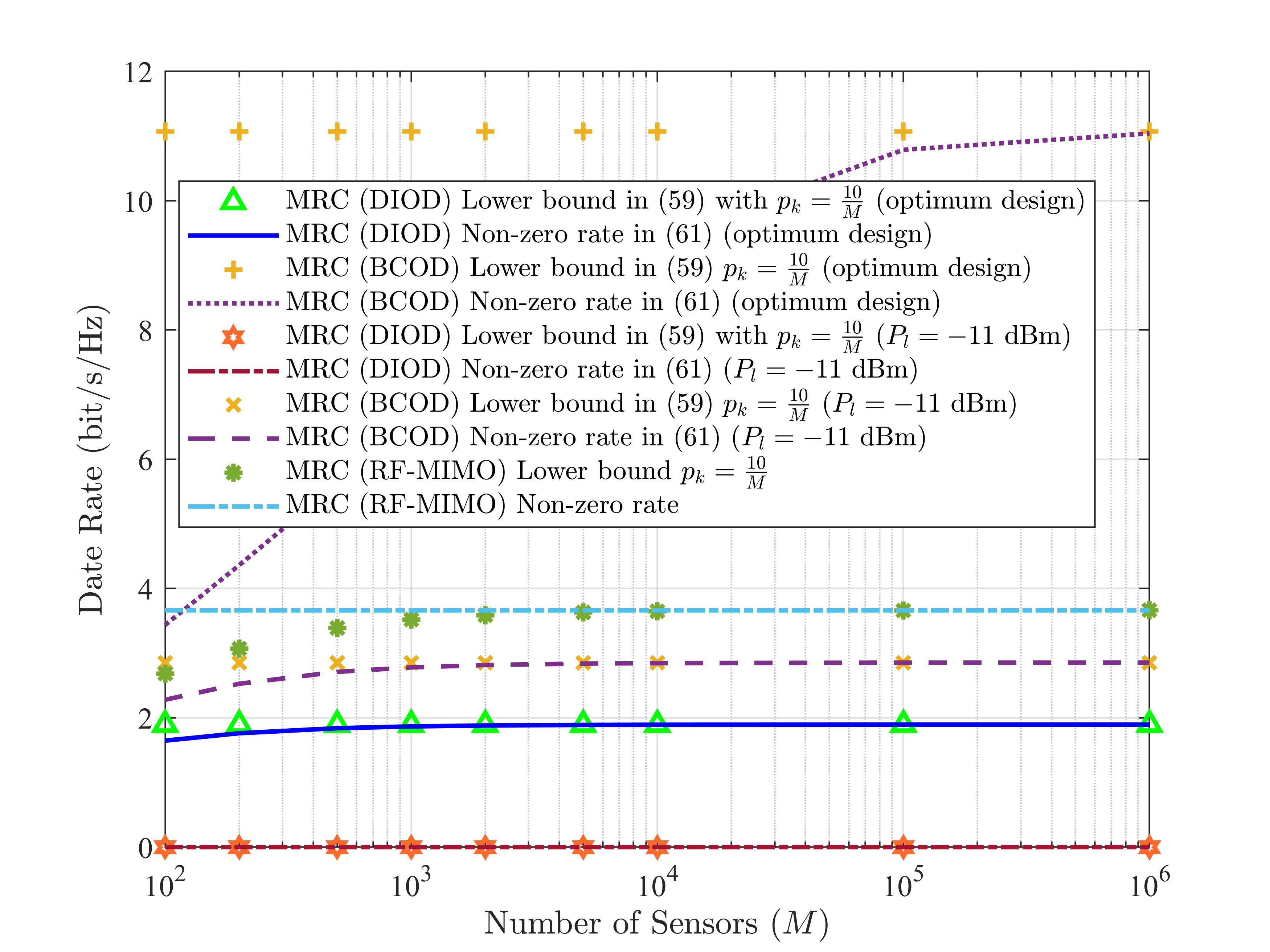}
        \subcaption{MRC detection.}
        \label{MRCpower}
    \end{minipage}\hfill
    \begin{minipage}[t]{0.45\linewidth}
        \centering
        \includegraphics[width=\linewidth]{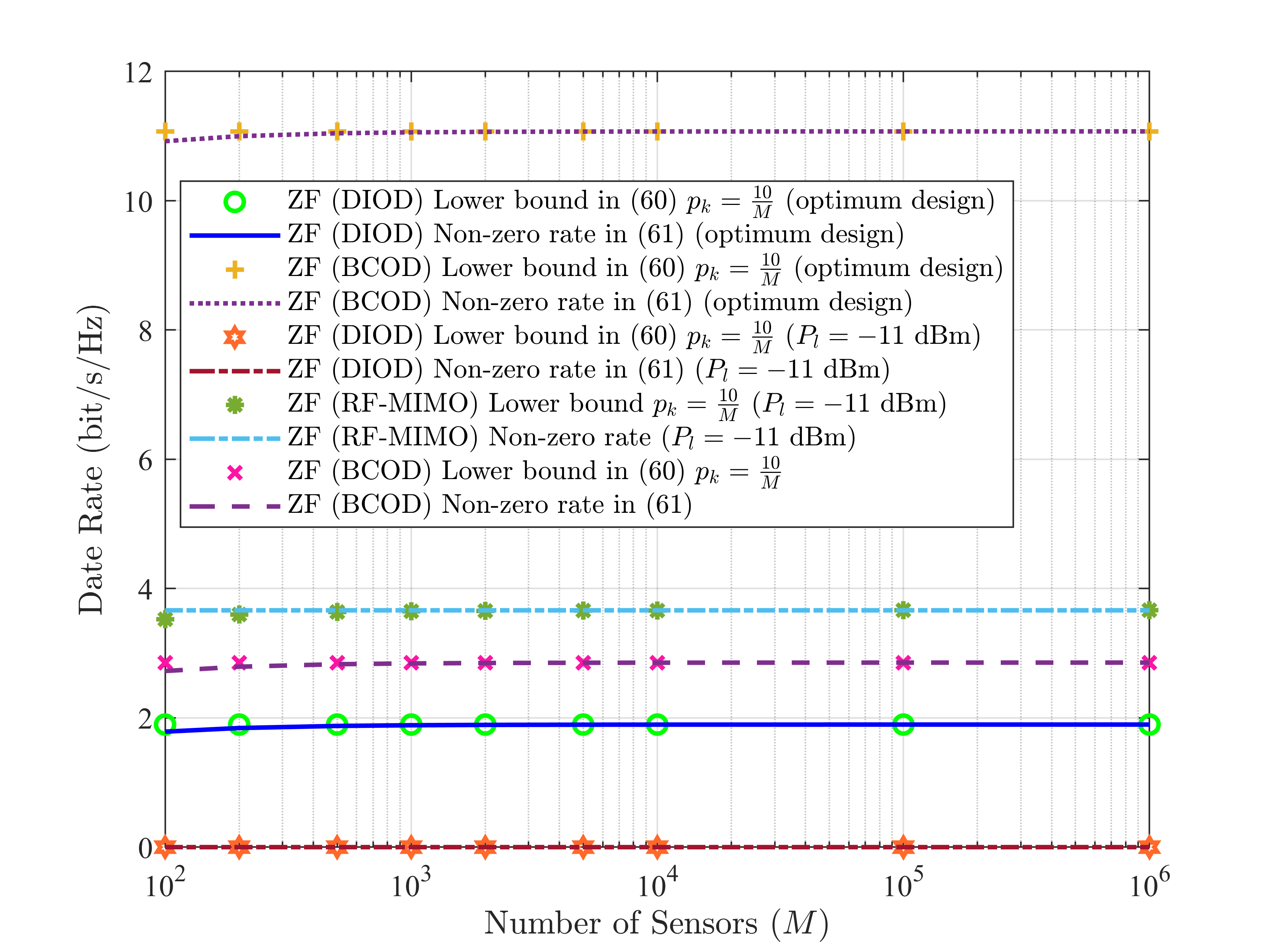}
         \subcaption{ZF detection.}
        \label{ZFpower}
    \end{minipage}\hfill

    \caption{\textcolor{black}{Power scaling law of RAQ-MIMO and RF MIMO  systems.}}
    \label{powerscaling}
     \vspace{-0.5cm}
\end{figure*}

\begin{figure*}[ht]
    \centering
    \begin{minipage}[t]{0.45\linewidth}
        \centering
        \includegraphics[width=\linewidth]{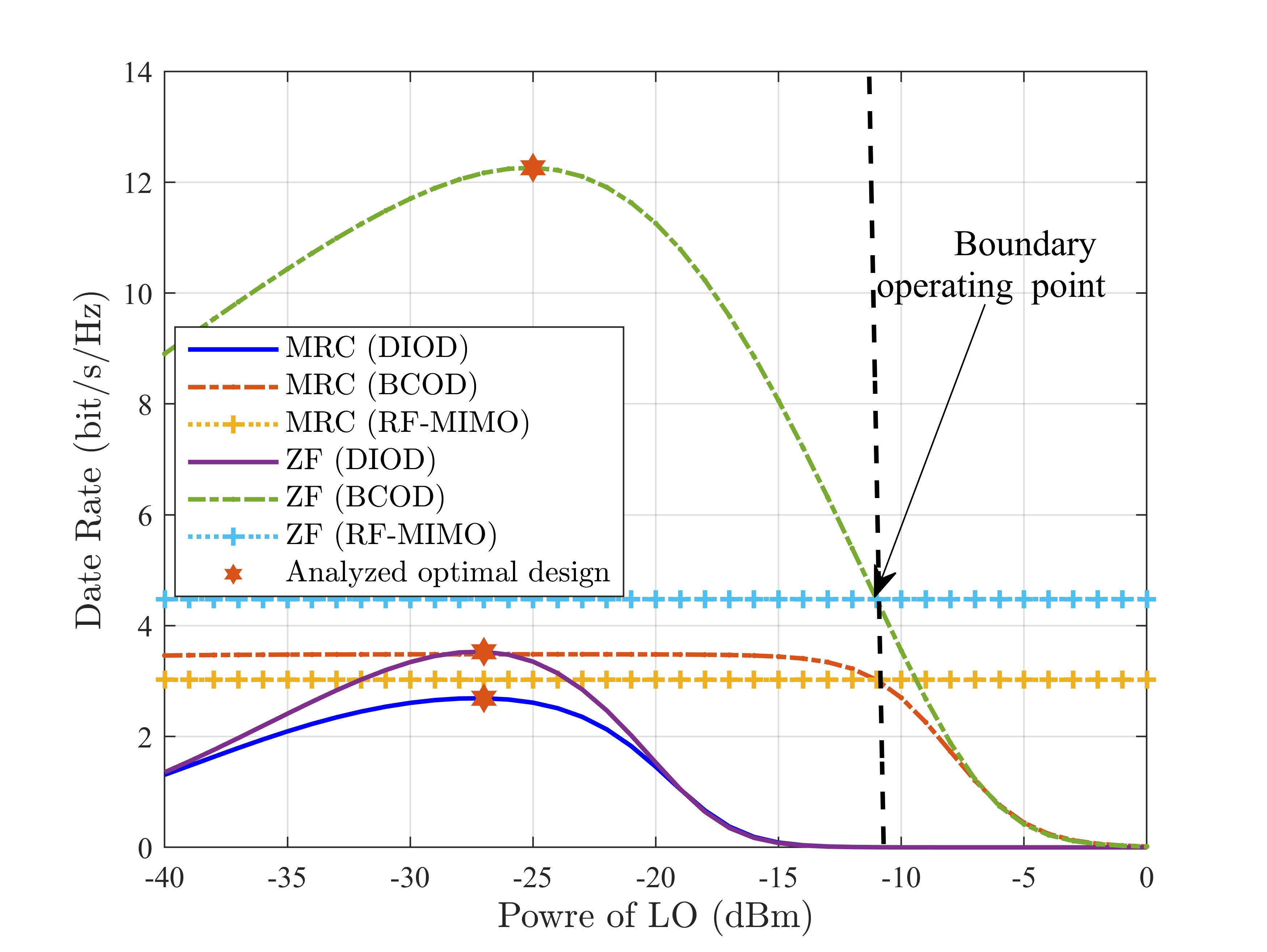}
        \subcaption{Data rate with various LO.} 
        \label{RAQLO}
    \end{minipage}\hfill
    \begin{minipage}[t]{0.45\linewidth}
        \centering
        \includegraphics[width=\linewidth]{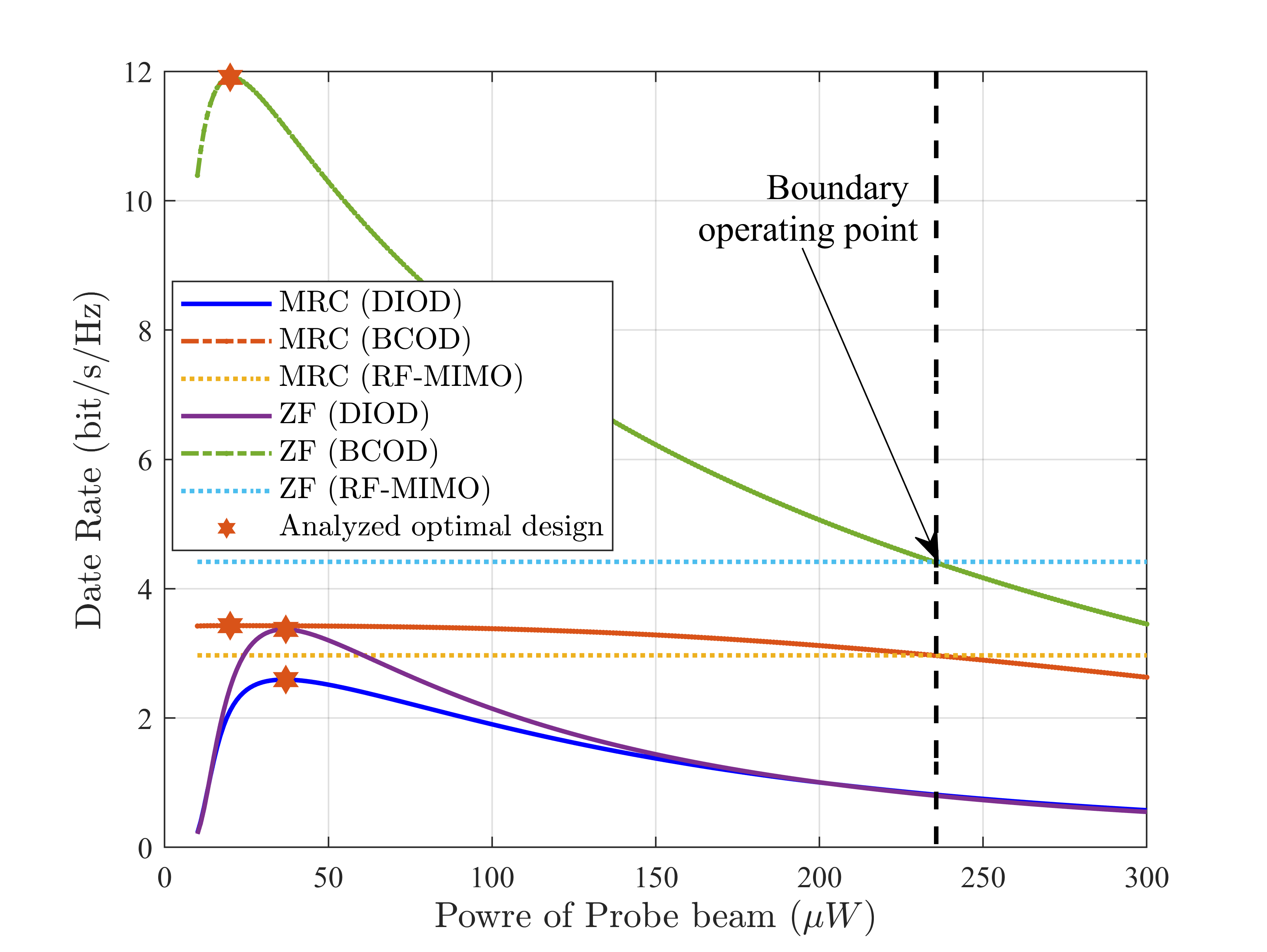}
         \subcaption{Data rate with various probe beam.}
        \label{RAQprobe}
    \end{minipage}\hfill

    \caption{Data rate of RAQ-MIMO with various system parameters. \textcolor{black}{The vertical dashed lines in (a) and (b) indicate the LO and probe-beam thresholds at which the inequality (63) holds with equality.}}
    \label{ratewithpara}

\end{figure*}

In this subsection, we check the waveform based on the master equation and our derived results with a fixed user location. Fig. \ref{Waveform} illustrates that the exact waveforms match the approximated ones and demonstrates that our derived model effectively captures the waveform variations of the received signal, thereby providing a traceable equivalent model for analysis and system design. By setting \(\varsigma^2 = 2qB\), we evaluate the gap between the signal-dependent shot noise for various values of \(\frac{U_\text{LO}}{U_x}\) in Fig. \ref{powerSN}. As shown, the approximated model closely matches the master-equation result whenever \(\frac{U_\text{LO}}{U_\text{x}}\ge 10 \) dB, numerically confirming the strong-LO regime assumed in Eq. (9). Furthermore, the signal-dependent shot noise of BCOD is larger than that of DIOD, since the noise level is highly dependent on the local optical power \(P_l\).

By setting \(\varsigma^2 = \textcolor{red}{2Bq}\), we investigate the impacts of system parameters on the received SNR under diverse noise sources in Fig. \ref{SISOTN}. As expected, the optimal SNR operating points under all individual noise sources agree well with the derived expressions, thereby validating the analysis and providing clear guidance for RAQR design. We further observe that DIOD is more sensitive to thermal noise, whereas BCOD is primarily limited by signal-dependent shot noise. Consequently, neglecting this noise source can lead to suboptimal system design. Finally, compared with the coupling beam, system performance is more strongly influenced by the LO and the probe beam.

Then, Fig. \ref{SISOloss} illustrates the performance loss between the operating design performance under ideal resonance and the optimal design performance under detuning conditions. The results show that the design derived for perfect resonance remains effective when the detuning is small. In addition, detuning has a more pronounced impact on DIOD, whereas BCOD is less sensitive because its performance is dominated by the local optical beam. More importantly, once signal-dependent shot noise is taken into account, the effect of detuning on the optimal operating point becomes significant.

\subsection{MIMO}

Building on the previously derived equivalent model, we generalize it to the multi-user setting by the theoretical projection and then evaluate the performance gap between our derived lower bounds and Monte-Carlo simulation. By setting \(K = 10\) and \(\varsigma^2 = \textcolor{red}{2qB}\), we depict the per-user data rate obtained from Monte Carlo simulations and the theoretical results in Fig.~\ref{ratevsM}. Similar to conventional RF antennas, increasing the number of atomic sensors can improve the data rate. Additionally, Monte Carlo simulations validate the accuracy of our derivations, providing a traceable lower bound for system analysis and design.

To further highlight the effects of different system parameters on RAQ-MIMO, we introduce conventional RF-MIMO as a baseline with the same array size \(M\), \(K\)-user setup, and large-scale fading model. First, Fig.~\ref{powerscaling} illustrates the power scaling law of RAQ-MIMO and RF MIMO systems. As shown, a non-zero achievable rate is maintained when the transmit power scales inversely with the number of atomic sensors, i.e., proportional to \(\frac{1}{M}\), validating our Remark 1.

\textcolor{black}{Fig.~\ref{ratewithpara} compares the lower bounds of RAQ-MIMO and RF-MIMO for \(M=100\). Consistent with our analysis, both the LO power and probe-beam power exert a highly non-monotonic impact on RAQ-MIMO, making its advantage conditional on proper front-end design. The vertical dashed lines in Fig.~\ref{ratewithpara}(a) and (b) graphically embody the crossover condition (63), marking the LO and probe-beam thresholds at which (63) holds with equality. Once these thresholds are crossed, the RAQ-MIMO rate degrades sharply and may even fall below that of RF-MIMO, confirming that RAQ-MIMO is not universally superior. Furthermore, thermal noise dominates the DIOD scheme, and thus its design should target the thermal-noise-limited regime. In contrast, signal-dependent shot noise dominates the BCOD scheme, where the design must balance the effective receive gain against the shot noise.}


\section{Conclusions}
This paper established a complex baseband equivalent model for photodetection-induced signal-dependent shot noise in superheterodyne RAQRs, providing a bridge between RF signal processing and atomic optical readout at the communication-theoretic level. The proposed model decomposed the shot noise into user-signal-dependent and direct-current-dependent components and showed that the optical operating point jointly set the normalized reception gain and the aggregate noise floor, thereby revealing an intrinsic gain–noise trade-off dictated by the system design. Building on this insight, the composite noise structure of RAQRs was systematically characterized, and explicit closed-form design criteria for practical optical front-end optimization were derived. By extending the analysis to RAQ-MIMO, we demonstrated that both the asymptotic achievable rate and the performance advantage over RF-MIMO were critically sensitive to the system configuration and detection scheme. Simulation results confirmed that judicious parameter design was not merely beneficial but essential for RAQ-MIMO to realize superior performance gains over conventional MIMO systems.

\vspace{-0.1cm}
\begin{appendices}	
\section{Proof of theorem \ref{MRC_SINR_T}}
\label{Prooftheorem1}
Owing to \(\mathbf{c}_k = \Phi\mathbf{D}\mathbf{h}_k\), we have
\begin{equation}
    \small 
    \mathbb{E}\{\sqrt{\rho p_k}\Phi\mathbf{c}^{H}_k\mathbf{D}\mathbf{h}_k\} = M\sqrt{\rho  p_k}|\Phi|^2 \beta_k.
\end{equation}

Next, the power of the leaked signal can be expressed as
\begin{equation}
    \small 
    \begin{split}
        |\text{Ls}_k|^2 &= \mathbb{E}{\{|\sqrt{\rho p_k}\Phi\mathbf{c}^{H}_k\mathbf{D}\mathbf{h}_k|^2\}} - |\text{Ds}_k|^2 \\
        & = \rho p_k |\Phi|^4 \mathbb{E}\{\mathbf{ h}^H_k\mathbf{h}_k\mathbf{h}^H_k \mathbf{ h}_k\}-|\text{Ds}_k|^2 \\
        & = M \rho p_k |\Phi|^4\beta^2_k. 
    \end{split}  
\end{equation}

The inter-user interference is given by
\begin{equation}
    \small 
    \begin{split}
        |\text{UI}_{k,k'}|^2 &= \mathbb{E}{\{|\sqrt{\rho p_{k'}}\Phi\mathbf{c}^{H}_k\mathbf{D}\mathbf{h}_{k'}|^2\}} = M \rho p^d_{k'} |\Phi|^4 \beta_{k'}\beta_k.
    \end{split}  
\end{equation}

The shot noise is expressed as
\begin{equation}
    \begin{split}
        &|\text{SN}_{k,k'}|^2 =  \mathbb{E}{\{|\sqrt{\rho_{\text{SN}} p_{k'}}\Phi_{\text{SN}}\mathbf{c}^{H}_k\mathbf{B}\mathbf{D}\mathbf{h}_{k'}|^2\}}\\
        =& \rho_{\text{SN}}p_k|\Phi|^2 \mathbb{E}\{\mathbf{h}^H_k\mathbf{D}^H\mathbf{B}\mathbf{D}\mathbf{h}_{k'}\mathbf{h}^H_{k'}\mathbf{D}^H\mathbf{B}^H\mathbf{D}\mathbf{h}_k\}\\
        =&\Bigg\{\begin{array}{cc}
        2M\varsigma^2\rho_{\text{SN}}p_k|\Phi|^2|\Phi_{\text{SN}}|^2\beta^2_k   ,&  k = k',\\
   M\varsigma^2\rho_{\text{SN}}p_{k'}|\Phi|^2|\Phi_{\text{SN}}|^2\beta_k\beta_{k'}     ,& k \neq k'.
        \end{array}
    \end{split}
\end{equation}

Similarly, the noise can be written as
\begin{equation}
    \small 
    \begin{split}
        |\text{N}_{k}|^2 &= \mathbb{E}{\{|\mathbf{c}^{H}_k\mathbf{w}|^2\}} = M|\Phi|^2\sigma^2 \beta_k.
    \end{split}  
\end{equation}

Finally, we complete this proof by substituting the above results into (\ref{MRC_LB_rate}).

\section{Proof of theorem \ref{ZF_SINR_T}}
\label{Prooftheorem2}
Owing to \(\mathbf{C} =\Phi\mathbf{D}\mathbf{ H} [(\Phi\mathbf{D}\mathbf{ H})^H\Phi\mathbf{D}\mathbf{ H}]^{-1}\), we have
\begin{equation}
    \small 
    \mathbb{E}\{\sqrt{\rho p_k}\Phi\mathbf{c}^{H}_k\mathbf{D}\mathbf{h}_k\} = \sqrt{\rho p_k}.
\end{equation}

Then, the leaked signal based on ZF is expressed as
\begin{equation}
    \small 
    \begin{split}
        |\text{Ls}_k|^2 &= \mathbb{E}{\{|\sqrt{\rho p_k}\Phi\mathbf{c}^{H}_k\mathbf{D}\mathbf{h}_k|^2\}} - |\text{Ds}_k|^2=0 \\
    \end{split}  
\end{equation}
Similarly, the inter-user interference \(|\text{UI}_{k,k'}|^2\) is 0. 

Before deriving shot noise, we define
\begin{equation}\label{eq:B_trace_split}
\bar b \triangleq \frac{1}{M}\mathrm{tr}(\mathbf{B}),\qquad
\widetilde{\mathbf{B}}\triangleq \mathbf{B}-\bar b\mathbf{I}_M,
\end{equation}
and thus we obtain $\mathrm{tr}(\widetilde{\mathbf{B}})=0$ and $\mathbf{B}=\bar b\mathbf{I}_M+\widetilde{\mathbf{B}}$. Then, we have
\begin{equation}\label{eq:split_main}
\begin{split}
    \mathbf{c}_k^H\mathbf{B}\mathbf{D}\mathbf{h}_{k'}
&= \bar b\,\mathbf{c}_k^H\mathbf{D}\mathbf{h}_{k'} + \mathbf{c}_k^H\widetilde{\mathbf{B}}\mathbf{D}\mathbf{h}_{k'}\\
&= \left\{\begin{array}{cc}
     \bar b +\mathbf{c}_k^H\widetilde{\mathbf{B}}\mathbf{D}\mathbf{h}_{k'}, &  k = k',\\
    \mathbf{c}_k^H\widetilde{\mathbf{B}}\mathbf{D}\mathbf{h}_{k'}, & k\neq k'.
\end{array}\right.
\end{split}
\end{equation}
Then, by using \(\mathrm{tr}(\widetilde{\mathbf{B}}\widetilde{\mathbf{B}}^H)
=\mathrm{tr}(\mathbf{B}\mathbf{B}^H)-\frac{|\mathrm{tr}(\mathbf{B})|^2}{M}\), \(\mathbb{E}\{\mathrm{tr}(\mathbf{B}\mathbf{B}^H)\}= \sum_{m=1}^M \mathbb{E}\{b_m^2\}=M\varsigma^2\),
and
\(\mathbb{E}\{|\mathrm{tr}(\mathbf{B})|^2\}= \mathrm{Var}\Big(\sum_{m=1}^M b_m\Big)=M\varsigma^2\), we have
\begin{equation}
    |\text{SN}_{k,k'}|^2 = \Bigg\{\begin{array}{cc}
       \rho_\text{SN}p_k\frac{|\Phi_\text{SN}|^2}{|\Phi|^2}\Big[ \frac{\varsigma^2}{M}
+\frac{\beta_{k'}}{\beta_k} \frac{(M-1)\varsigma^2}{M(M-K)}\Big] ,  & k = k',  \\
        \rho_\text{SN}p_{k'}\frac{|\Phi_\text{SN}|^2}{|\Phi|^2} \Big[ \frac{\beta_{k'}}{\beta_k} \frac{(M-1)\varsigma^2}{M(M-K)}\Big] , & k \neq k'.
    \end{array}
\end{equation}
\textcolor{black}{Note that this result only holds for \(M > K\). For \(K = M\), this result diverges.}

The noise is expressed as
\begin{equation}
    \small 
    \begin{split}
        |\text{N}_{k}|^2 &= \mathbb{E}{\{|\mathbf{c}^{H}_k\mathbf{w}|^2\}}= \frac{\sigma^2 }{(M-K)|\Phi|^2 \beta_k}.
    \end{split}  
\end{equation}

By combining the results, the proof of Theorem \ref{ZF_SINR_T} is completed.

\end{appendices}	


\bibliographystyle{IEEEtran}
\bibliography{myref}

\end{document}